\documentclass[iop]{emulateapj}
\usepackage{natbib}
\usepackage{comment}
\usepackage{xcolor}
\usepackage{longtable}
\usepackage{savesym}
\savesymbol{tablenum}
\usepackage{siunitx}
\restoresymbol{SIX}{tablenum}

\begin{document}
\title{EVR-CB-001: An Evolving, Progenitor, White Dwarf Compact Binary \\Discovered with the Evryscope}

\author{Jeffrey~K.~Ratzloff\altaffilmark{1}, Brad~N.~Barlow\altaffilmark{2}, Thomas~Kupfer\altaffilmark{3}, Kyle~A.~Corcoran\altaffilmark{2,4}, Stephan~Geier\altaffilmark{5}\\ Evan~Bauer\altaffilmark{6}, Henry~T.~Corbett\altaffilmark{1}, Ward~S.~Howard\altaffilmark{1}, Amy~Glazier\altaffilmark{1}, and Nicholas~M.~Law\altaffilmark{1}}

\altaffiltext{1}{Department of Physics and Astronomy, University of North Carolina at Chapel Hill, Chapel Hill, NC 27599-3255, USA}
\altaffiltext{2}{Department of Physics and Astronomy, High Point University, High Point, NC 27268, USA}
\altaffiltext{3}{Kavli Institute for Theoretical Physics, University of California, Santa Barbara, CA 93106, USA}
\altaffiltext{4}{Department of Astronomy, University of Virginia, Charlottesville, VA 22904, USA}
\altaffiltext{5}{Institut für Physik und Astronomie, Universit\"at Potsdam, Haus 28, Karl-Liebknecht-Str. 24/25, 14476, Potsdam-Golm, Germany}
\altaffiltext{6}{Department of Physics, University of California, Santa Barbara, CA 93106, USA}

\email[$\star$~E-mail:~]{jeff215@live.unc.edu}


\begin{abstract}

We present EVR-CB-001, the discovery of a compact binary with an extremely low mass ($.21 \pm 0.05 M_{\odot}$) helium core white dwarf progenitor (pre-He WD) and an unseen low mass ($.32 \pm 0.06 M_{\odot}$) helium white dwarf (He WD) companion. He WDs are thought to evolve from the remnant helium-rich core of a main-sequence star stripped during the giant phase by a close companion. Low mass He WDs are exotic objects (only about .2$\%$ of WDs are thought to be less than .3 $M_{\odot}$), and are expected to be found in compact binaries. Pre-He WDs are even rarer, and occupy the intermediate phase after the core is stripped, but before the star becomes a fully degenerate WD and with a larger radius ($\approx .2 R_{\odot}$) than a typical WD. The primary component of EVR-CB-001 (the pre-He WD) was originally thought to be a hot subdwarf (sdB) star from its blue color and under-luminous magnitude, characteristic of sdBs. The mass, temperature ($T_{\rm eff}=18,500 \pm 500 K$), and surface gravity ($\log(g)=4.96 \pm 0.04$) solutions from this work are lower than values for typical hot subdwarfs. The primary is likely to be a post-RGB, pre-He WD contracting into a He WD, and at a stage that places it nearest to sdBs on color-magnitude and $T_{\rm eff}$-$\log(g)$ diagrams. EVR-CB-001 is expected to evolve into a fully double degenerate, compact system that should spin down and potentially evolve into a single hot subdwarf star. Single hot subdwarfs are observed, but progenitor systems have been elusive.


\end{abstract}


\section{INTRODUCTION} \label{section_intro}

Compact binaries are highly sought after and studied objects because of their potential to test stellar formation and evolution theory, and measure primary and secondary parameters to high precision. Compact binaries are the suspected progenitors to astrophysical phenomena that are not well understood, including many classes of supernovae, single hot subdwarf B (sdB) stars, and low-mass white dwarfs (see \citealt{2014LRR....17....3P} and references therein). The primary and secondary components of compact binaries influence the usefulness of the system to explain key formation or evolution phases \citep{1995MNRAS.275..828M}. Photometric variability from eclipses, ellipsoidal deformation, gravitational limb darkening, Doppler beaming, or from combinations of these effects enables system parameters to be solved more fully and with higher precision than from radial velocity alone \citep{2016MNRAS.458..845H, 2013A&A...554A..54G, 2011MNRAS.410.1787B}. Systems that are evolving into highly-sought-after and poorly understood objects are useful for testing theory and allowing for detailed observations of rare progenitors \citep{2011MNRAS.413.1121R, 2041-8205-759-1-L25, 2015AJ....149..176D, 2005A&A...440.1087N}. We discuss these points in the context of white dwarf (WD) binaries and show EVR-CB-001 to be a rare combination of almost all of these desired traits.

Many compact binaries are thought to form as stars evolve from the main sequence to the giant phase, with an increasing radius that engulfs the companion and facilitates mass transfer. If the companion is unable to accrete at a high enough rate, a common envelope (CE) is formed and angular momentum is transferred to the envelope, thereby decreasing the orbital period. Eventual ejection of the CE leaves behind a compact binary with short orbital period \citep{2015A&A...576A..44K}. Double WD or WD / sdB binaries with orbital periods below a few hours lose angular momentum predominantly from gravitational wave radiation, meaning that once the CE phase is completed this type of compact binary system will remain relatively unchanged. These systems are good candidates to study the CE phase, especially the later stage \citep{1995MNRAS.275..828M}. EVR-CB-001 is a WD binary of this type with a potentially clean post CE phase.

The CE phase is also important for understanding the formation processes that lead to sdBs and low-mass He WDs. Low mass WDs must be formed through binary interactions (e.g. \citealt{1995MNRAS.275..828M}) and most hot subdwarfs are thought to form from a red giant progenitor that is stripped of its outer hydrogen envelope during CE interactions with a nearby companion \citep{han02, han03}. This process leaves behind a $\approx$0.5$M_{\odot}$ helium-burning core (the hot subdwarf) in a close orbit with the companion that led to its formation. A comprehensive summary of hot subdwarfs can be found in \cite{heb86,heb09,2016PASP..128h2001H}. If the mass of the He core in the progenitor is not high enough to start He burning when the star gets stripped the object will bypass the horizontal branch and contract onto the white dwarf cooling sequence as a He WD \citep{ist16, 2010ApJ...723.1072B}. If the He core is relatively young (not long past the CE phase), then it will be a pre-He WD with similar spectroscopic characteristics to an sdB (temperature, color and absolute magnitude) but with lower mass. Discovery of a pre-He WD at this juncture offers an opportunity to study a key intermediate stage of the He WD. The EVR-CB-001 primary is a pre-He WD apparently caught at a very early stage, with sdB--like characteristics; we actually discovered EVR-CB-001 in an sdB variability search due to its similar color/magnitudes.

The combined mass and the mass ratio of the primary and the companion is a key driver for studying compact binaries. Compact binary searches originally targeted high mass WD/WD or WD/sdB mergers as they are thought to be the most promising type Ia supernovae (SN Ia) progenitors \citep{2012NewAR..56..122W}. Compact binaries with the necessary mass and short period have proven elusive, with just a handful of promising candidates despite decades of searching \citep{2013A&A...554A..54G,2041-8205-759-1-L25, 2000MNRAS.317L..41M, gei07} and \citep{2015AJ....149..176D, 2005A&A...440.1087N}.

More recently, searches have aimed to find low-mass systems with the goal of explaining the formation of He WDs and other exotic objects. He WDs with masses less than $.3M_{\odot}$ do not have a known mechanism to fuse helium, and would have to evolve from the giant phase and cool to form final He WD. This is expected to take longer than the age of the galaxy \citep{1995MNRAS.275..828M}. A CE stage from a close companion would interrupt this lengthy stellar evolution process, and extremely low mass He WDs are expected to be members of compact binaries. The ELM project, using color-color cuts from the SDSS and spectra from the Hypervelocity Star Survey (HVS), has found a few dozen extremely low mass He WDs as well as pre-He WDs \citep{2016ApJ...818..155B, 2014ApJ...794...35G, 2012ApJ...751..141K, 2010ApJ...723.1072B}.

Compared to these known extremely low mass He WD systems, the primary of EVR-CB-001 has the lowest surface gravity of all known systems and a higher temperature than all but a few, and is quite rare in that the primary and secondary are both extremely low mass for WDs. The system is compact with a fast period, and will evolve into a fully double degenerate binary. It is then expected to shrink via gravitational wave radiation, and merge into a single helium-rich object or if the merge can be prevented into a stably accreting AM\,CVn binary. In \S~\ref{section_discussion} we discuss EVR-CB-001 as viable progenitor candidate for a single hot subdwarf with a mass (estimated from the pre-merger mass of $.47M_{\odot}$), very close to the canonical hot subdwarf mass.

Only a small fraction of compact detached binaries show photometric ellipsoidal and radial velocity variations necessary for detailed solutions. Only five such fast-period hot subdwarf $+$ WD binaries have been published in the literature \citep{1998MNRAS.300..695K, 2000MNRAS.317L..41M, 2041-8205-759-1-L25, Kupfer_2017, 2017ApJ...851...28K}, and fewer than ten WD/WD compact systems with either eclipses or ellipsoidal modulations \citep{2016MNRAS.458..845H}. EVR-CB-001 shows high amplitude photometric variability with multiple components, large radial velocity variations, and it is bright ($m_{G} = 12.581 \pm .003$), characteristics which allow for a precise solution of the system.

Here we report the discovery of the pre-He WD$+$He WD binary Gaia DR2 5216785445160303744 (hereafter, ``EVR-CB-001''), which shows strong ellipsoidal modulations and gravitational darkening. We note that ASAS-SN listed the source as a unidentifiable variable (ASASSN-V J084815.55-741854.3) in \citep{2019MNRAS.485..961J}. EVR-CB-001 was found from a southern all-sky hot subdwarf survey searching for low-mass companions (Ratzloff et al., in prep) using the Evryscope.

This paper is organized as follows: in \S~\ref{section_obs_phot} we describe the observations and reduction. In \S~\ref{section_analysis_spectra} we describe our spectroscopic analysis to determine the orbital and atmospheric parameters of the pre-He WD. In \S~\ref{section_analysis_lc} we model the photometric light curve to determine ellipsoidal modulations and test for eclipses. In \S~\ref{section_system_solution} we solve the system and show our results. In \S~\ref{section_discussion} we discuss our findings and conclude in \S~\ref{section_summary}.


\section{Observations \& Reduction} \label{section_obs_phot}

\subsection{Evryscope Photometry}

We discovered photometric oscillations in EVR-CB-001 from analyzing 2.5 years of data from the Evryscope, obtained from January, 2016 to June, 2018. Data were taken through  a Sloan {\em g} filter with 120 s integration times, providing a total of 53,698 measurements. The wide-seeing Evryscope is a gigapixel-scale, all-sky observing telescope that provides new opportunities for uncovering rare compact binaries through photometric variations. It is optimized for short-timescale observations with continuous all sky coverage and a multi-year period observation strategy. The Evryscope is a robotic camera array mounted into a 6 ft-diameter hemisphere which tracks the sky \citep{2015PASP..127..234L, 2019arXiv190411991R}. The instrument is located at CTIO in Chile and observes continuously, covering 8150 sq. deg. in each 120s exposure. Each camera features a 29MPix CCD providing a plate scale of 13"/pixel. The Evryscope monitors the entire accessible Southern sky at 2-minute cadence, and the Evryscope database includes tens of thousands of epochs on 16 million sources.

Here we only briefly describe the calibration, reduction, and extraction of light curves from the Evryscope; for further details we point the reader to our Evryscope instrumentation paper \citep{2019arXiv190411991R}. Raw images are filtered with a quality check, calibrated with master flats and master darks, and have large-scale backgrounds removed using the custom Evryscope pipeline. Forced photometry is performed using APASS-DR9 \citep{2015AAS...22533616H} as our master reference catalog. Aperture photometry is performed on all sources using multiple aperture sizes; the final aperture for each source is chosen to minimize light curve scatter. Systematics removal is performed with a custom implementation of the SysRem \citep{2005MNRAS.356.1466T} algorithm.

\begin{figure}[t]
\includegraphics[width=1.0\columnwidth]{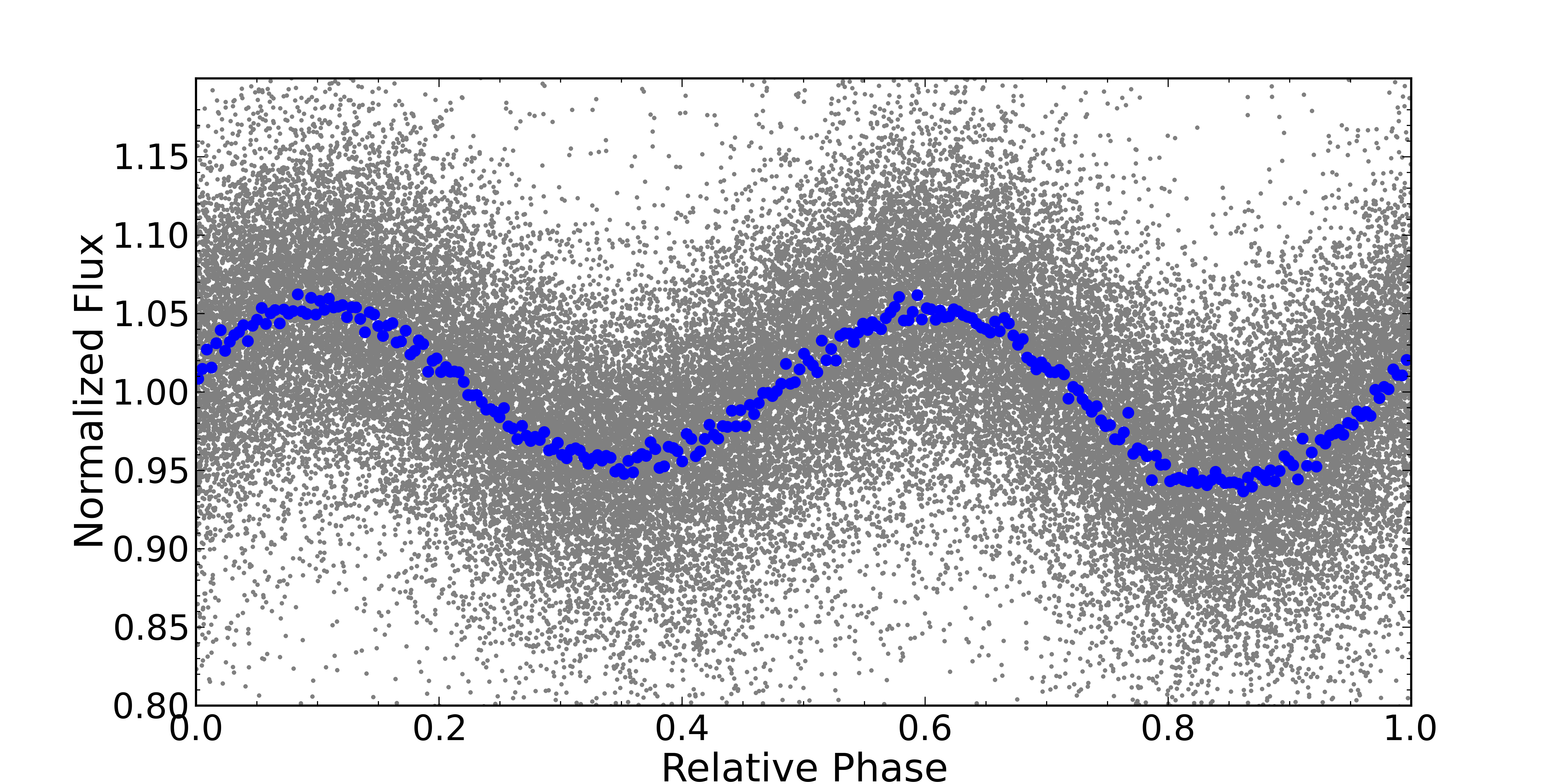}
\includegraphics[width=1.0\columnwidth]{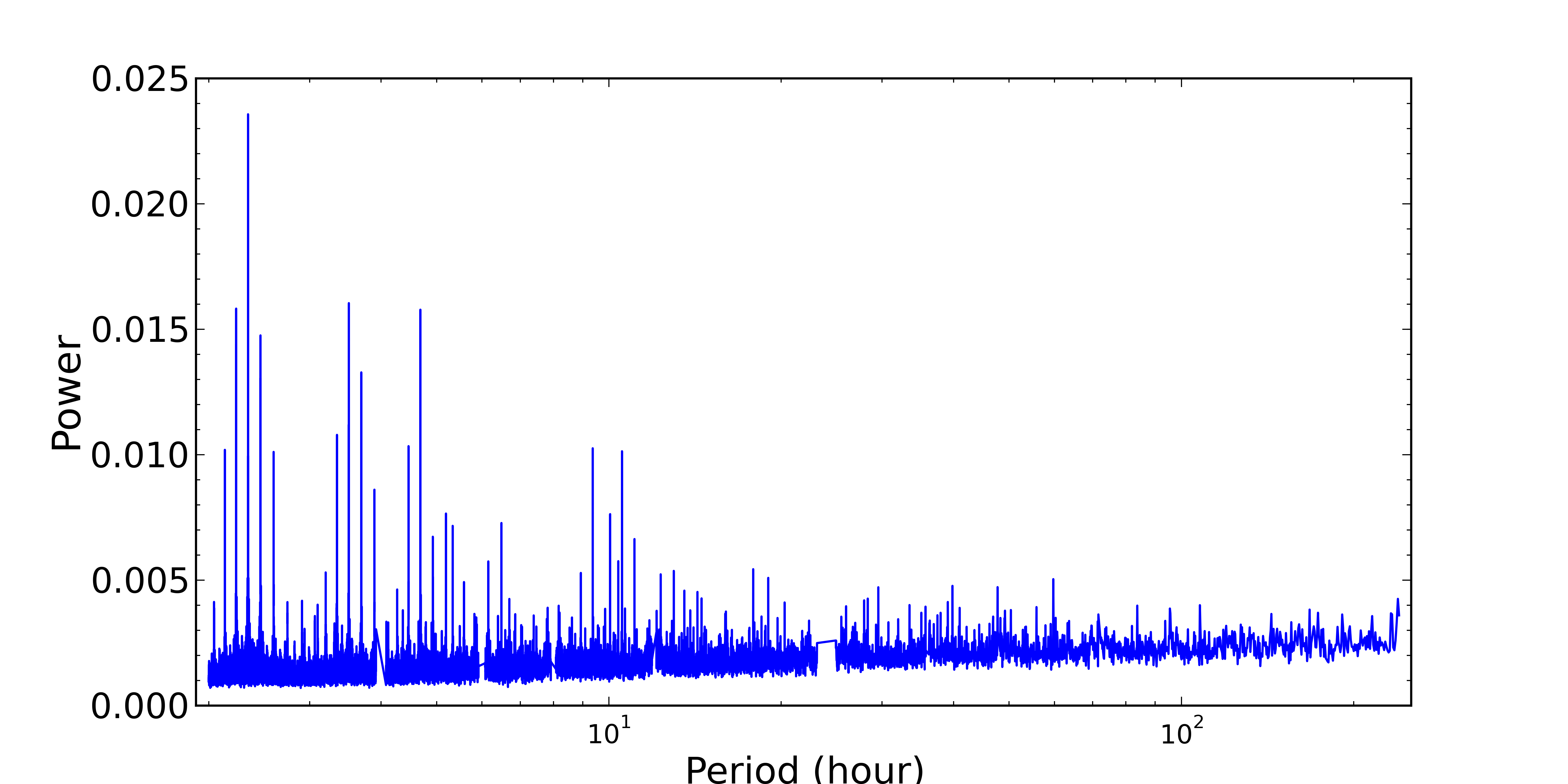}
\caption{The Evryscope discovery light curve of EVR-CB-001 folded on its period of 2.34249 hours is shown on the top panel. Grey points = 2 minute cadence, blue points = binned in phase. The bottom panel shows the BLS power spectrum with the highest peak at the 2.34249 hour detection.}
\label{fig:discovery}
\end{figure}

We use a panel-detection plot that filters the light curves, identifies prominent systematics, searches a range of periods, and phase folds the best detections from several algorithms for visual inspection. It includes several matched filters to identify candidate hot subdwarfs for variability and is described in detail in \citep{polarpaper}. EVR-CB-001 was discovered using Box Least Squares (BLS; \citealt{Kovacs:2002gn, 2014A&A...561A.138O}) with the same settings, pre-filtering, and daily-alias masking described in \citet{polarpaper}. The discovery tools and settings were tested extensively to maximize recovery of the fast transits and eclipses characteristic of hot subdwarfs and white dwarfs. As part of our testing, we also recovered CD-30 \citep{2041-8205-759-1-L25}, the only known fast-period hot subdwarf $+$ WD binary in our field of view and magnitude range. The BLS power spectrum revealed EVR-CB-001 to be a 2.34 hr binary exhibiting strong (12\%) modulations due to the ellipsoidal deformation of the primary from the unseen, more massive companion. The detection power in terms of Signal Detection Efficiency (SDE) \citep{Kovacs:2002gn} is 33.5, compared to an average SDE of 8 for targets in the hot subdwarf survey (Ratzloff et al., in prep) that EVR-CB-001 was discovered in. Figure \ref{fig:discovery} presents both the BLS power spectrum and phase-folded light curve.

Our detection tools also include Lomb-Scargle (LS) \citep{1975Ap&SS..39..447L, 1982Ap&SS..263..835S} and interestingly, the LS detection of the short periods in both EVR-CB-001 and CD-30 are relatively weak and are overpowered by longer periods (the search range in our survey is 2-720 hours for LS in an effort to recover a wide range of variables). Narrowing the period search range and further filtering of low frequency signals recovers the same period from LS as the BLS discovery. The high amplitude photometric variability in EVR-CB-001 results in an asymmetric signal, with a difference in even versus odd phase that is significant enough to affect the LS (optimized for sinusoidal signals) recovery. In this work, we show the BLS signal as it is the algorithm that led to the discovery, and is confirmed to be the correct period in \S~\ref{section_analysis_spectra}. In the Evryscope hot subdwarf survey (Ratzloff et al., in prep), we compare the effectiveness of BLS and LS in detecting compact ellipsoidal systems with multiple light curve features and discuss the modifications to our original search in an effort to maximize the recovery of these compact binary systems.

A subtle asymmetry in the light curve (a sub 1\% difference in the height of alternating peaks) is observed, indicative of Doppler boosting with the higher peak corresponding to the orbital position where the pre-He WD is moving toward us most quickly. The difference in minima is due to gravitational darkening of the deformed primary, with the lower minimum corresponding to the orbital position where the pre-He WD is farthest from us. 

\subsection{SOAR/Goodman Photometry}\label{section_soar_photometry}

In order to obtain a higher S/N light curve for modeling, we observed EVR-CB-001 on January 5, 2019 using the SOAR 4.1 m telescope at Cerro Pachon, Chile, with the Goodman spectrograph \citep{2004SPIE.5492..331C} in imaging mode. We used the blue camera with Bessel-V blocking filter and took 409 images with 15 s integration times. The image Region of Interest (ROI) was reduced to 1700 x 1000 pixels with 1x1 binning, which resulted in a 60\% duty cycle. The surrounding field is sparse, and so a larger-than-ideal ROI was needed to capture a sufficient number of comparison stars. For calibrations, we took 10 dome flats using 25\% lamp power and 10 s integrations, 10 darks also with 10 s integrations, and 10 bias frames.

The SOAR frames were processed with a custom aperture photometry pipeline written in Python. The object images were bias-subtracted, dark-subtracted, and flat-field-corrected using master calibration frames. Five reference stars of similar magnitude were selected, and aperture photometry was performed on all frames using a centroid algorithm and range of aperture sizes. The reference stars were confirmed to be non-variable. We also use the photometric aperture on dark areas of the image near the reference stars to capture background counts. The reference star counts are combined for the image and the background is subtracted (using the average per-pixel background times the pixels in the aperture). The background subtracted reference star counts are recorded for each image, and normalized by the mean. The target star counts are background subtracted in the same way and recorded for each image. The background subtracted target star counts are divided by the normalized background subtracted reference star counts to remove sky variations. The result is normalized by the mean to produce the final light curve.

In order to choose the best aperture, we removed variability from each light curve and chose the aperture with the lowest residual rms values. At this juncture, we did not have an exact model of the astrophysical variability of the system, but needed a reliable estimate of the variability so that it could be removed to measure the residual rms and choose the best photometric aperture. We used a Savitzk-Golay filter from the scipy.signal module \citep{scipy}, holding the the filter settings constant for all apertures. We also explored different settings to confirm the filter was not biasing the results. The filter was only used in this step to determine the best aperture, and is in no way applied to the photometry. The solution converged nicely with the minimum rms corresponding to a photometric aperture of 36 pixels, as shown in Figure \ref{fig:rms_aperture} in the appendix. The resulting differential light curve from SOAR, which we use to model EVR-CB-001, is shown later in the manuscript, in Figure \ref{fig:LC_SOAR}.

\begin{table*}[t]
\caption{Overview of Observations for EVR-CB-001}
\centering
\begin{tabular}{l l l l l}
\hline
Telescope & Date & Filter/Resolution & Epochs & Exposure\\
\hline
Photometry & & & &\\
Evryscope & Jan 2016 - Jun 2018 & Sloan \textit{g} & 53,698 & 2 min\\
SOAR/Goodman & Jan 5, 2019 & Bessel-V & 409 & 15 s\\
\hline
Spectroscopy & & & &\\
SMARTS 1.5-m/CHIRON & Dec 2018 - Jan 2019 & 28,000 & 29 & 600 s\\
SOAR/Goodman & Dec 2, 2018 & 1150 & 4 & 360 s\\
\hline
\end{tabular}
\label{observations_table}
\end{table*}

\subsection{SMARTS 1.5-m/CHIRON Spectroscopy}

We observed EVR-CB-001 on 29 nights between December 19, 2018 and January 28, 2019 with the SMARTS 1.5 m telescope and CHIRON, a fiber-fed cross-dispersed echelle spectrometer \citep{2013PASP..125.1336T}. Spectra were taken in image fiber mode (R $\sim$ 28000) and covered the wavelength range 4400-8800 \AA. We used integration times of 600 s to obtain just enough S/N for radial velocity measurements; longer integrations would have resulted in too much phase-smearing. Spectra were obtained every few days at specified epochs until full phase coverage was achieved. All raw spectra were reduced and wavelength-calibrated by the official CHIRON pipeline, housed at Georgia State University and managed by the SMARTS Consortium\footnote{http://www.astro.yale.edu/smarts/}. In addition to  H$\alpha$ and H$\beta$, which span multiple orders, the spectra show four He\,{\sc i} lines, including 6678 \AA, 5876 \AA, 5016 \AA, and 4922 \AA. All of these lines are synced in phase, with no signs of absorption due to a companion, and we conclude they emanate from a single star.

\subsection{SOAR/Goodman Spectroscopy}

The CHIRON spectra have too high a resolution to easily model atmospheric parameters using the H Balmer lines, which span multiple orders. As such, we also obtained low-resolution spectra on December 2, 2018 with the Goodman spectrograph using the 600 mm$^{-1}$ grating blue preset mode, 2x2 binning, and the 1" slit. This configuration provided a wavelength coverage of 3500-6000 \AA\ with spectral resolution of 4.3 \AA\ (R$\sim$1150 at 5000 \AA). We took four 360 s spectra of both the target and the spectrophotometric standard star BPM 16274. For calibrations, we obtained 3 x 60 s FeAr lamps, 10 internal quartz flats using 50\% quartz power and 30 s integrations, and 10 bias frames.

\begin{figure}[t]
\centering
\includegraphics[width=0.865\columnwidth]{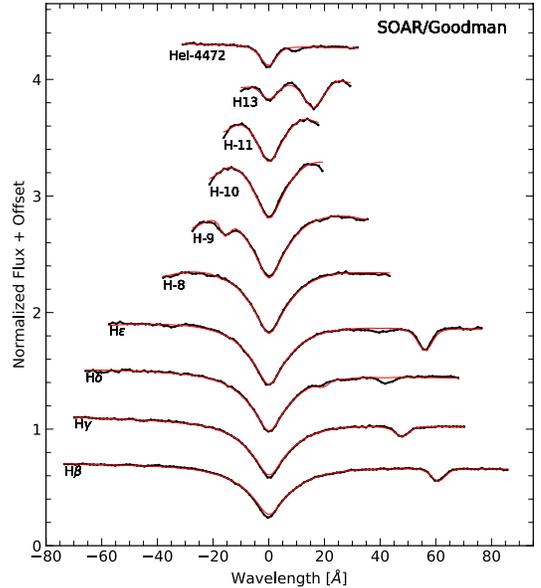}
\caption{Normalized SOAR/Goodman spectrum of EVR-CB-001 (black line) with best--fitting atmospheric model (red line). Parameters associated with the best-fitting LTE model spectrum are shown in the figure.}
\label{fig:spec_SOAR}
\end{figure}

We processed the spectra with a custom pipeline written in Python. The spectra were individually bias-subtracted and flat-corrected. A 3rd-order polynomial was fitted to the brightest pixels in each row; the spectra are then extracted in a 10-pixel range and background subtracted. We identify 16 prominent lamp emission lines and compare with the known lines of the FeAr lamp using a Gaussian fit to each feature. We used a 4th-order polynomial to fit the wavelength solution and calibrate each spectrum. We used our observations of BPM 16274 to flux-calibrate the EVR-CB-001 spectra by removing prominent absorption features and fitting a 7th-order polynomial to the continuum. Each spectrum was then rest-wavelength calibrated using a Gaussian fit to the H$\beta$ through H11 absorption features, as well as several prominent He absorption features. The resulting spectra were median-combined to form a final spectrum for atmospheric modeling. As shown in Figure \ref{fig:spec_SOAR}, we detect strong H Balmer lines, from H$\beta$ through H13, and one He\,{\sc i} line at 4472 \AA. As was the case for the CHIRON spectra, we find no evidence of absorption features due to the companion star; EVR-CB-001 appears to be a single-lined binary. Table \ref{observations_table} presents a brief overview of all of the photometric and spectroscopic data used in our analysis of EVR-CB-001. 


\section{Orbital and Atmospheric Parameters} \label{section_analysis_spectra}

The long baseline and dense coverage of the Evryscope photometry means we can determine the orbital period with high precision through O--C analysis. First, we converted all Evryscope time stamps from Modified Julian dates to Barycentric Julian dates, $BJD_{\mathrm{TDB}}$, using the web tool provided by \citet{Eastman10}. As an initial guess for the ephemeris, we used a Lomb Scargle periodogram to approximate the orbital period ($P$) and used a sine wave fit to the entire data set to estimate a reference time of minimum ($T_0$). From these, we generated several predicted times of minima (C values). Observed times of minima (O values) were determined by breaking up the entire Evryscope light curve into several segments, each containing approximately 10 orbits of data, and performing least-squares fits of sine waves to the segments. We then plotted O--C against O and adjusted $T_0$ and $P$ iteratively until there was no residual slope and the mean O--C value was zero. From this process we report the following orbital ephemeris for times of light minima, with $E$ representing the cycle number:

\begin{eqnarray*}
    t_{\mathrm min} &= \mathrm{BJD \, UTC} \,\, (2457812.75378 \pm 0.00005)\\
   &\phantom{=}+ (0.09760507 \pm 0.00000002 \, \mathrm{d})\times E
\end{eqnarray*}

\noindent The O-C diagram is fitted well with a linear trend, and we currently find no statistically significant evidence of a parabolic trend due to secular evolution or oscillations from reflex motion. We limit changes in the orbital period to $|\dot{P}| < 8 \times 10^{-9}$ s s$^{-1}$.

\begin{figure}[h]
\centering
\includegraphics[width=1\columnwidth]{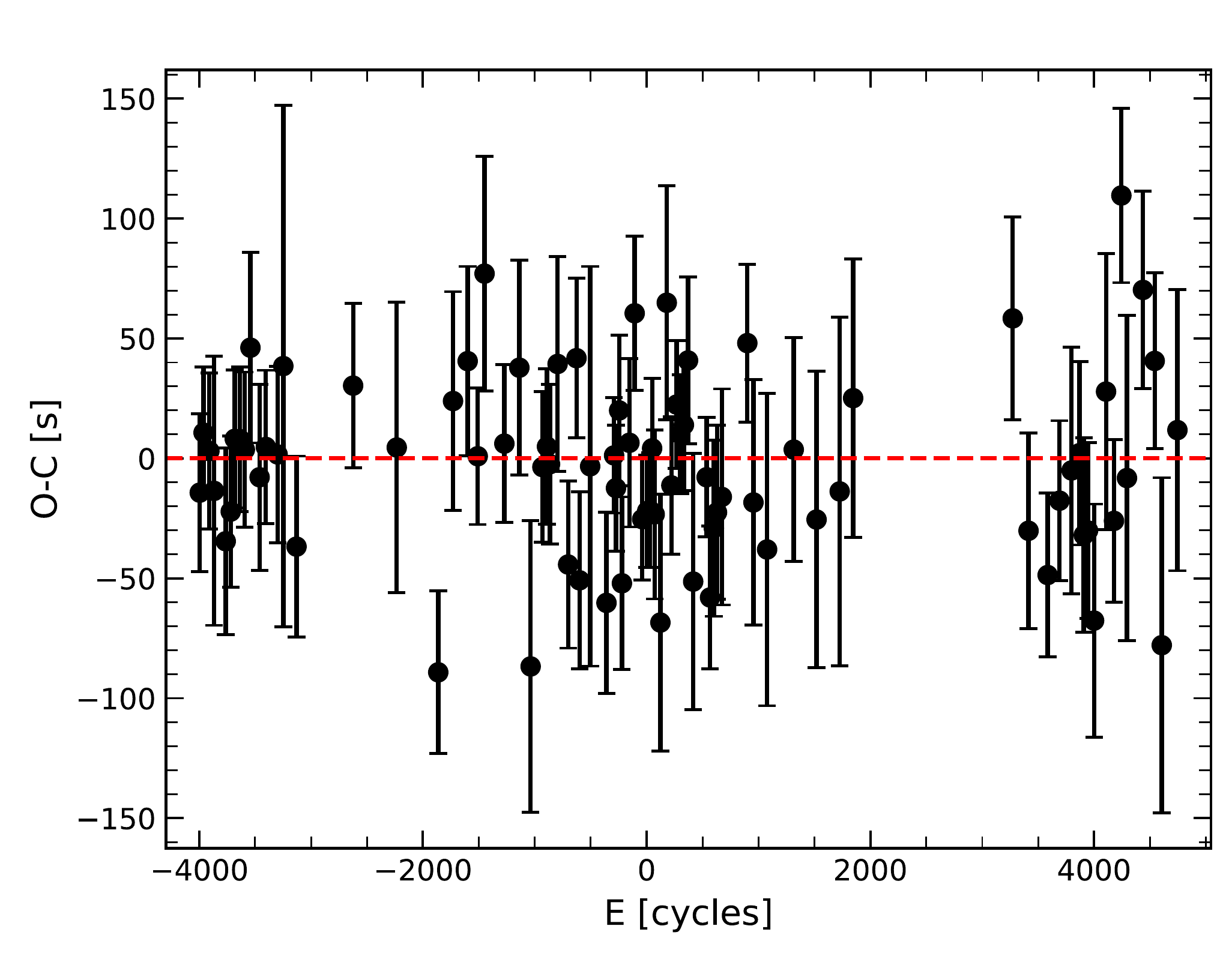}
\caption{O--C diagram constructed from the Evryscope light curve. The 2.5--year light curve was broken into 77 segments, each with 10 orbits worth of data ($\sim$700 measurements), and sine waves were fitted to the segments to determine phases. We limit any changes in the orbital period to $|\dot{P}| < 8 \times 10^{-9}$ s s$^{-1}$.}
\label{fig:EVR001_fig}
\end{figure}

Radial velocities were determined using data from CHIRON. We visually inspected each spectral order and chose the following high signal-to-noise absorption features for fitting: He\,{\sc i} 4922 \AA, He\,{\sc i} 5016 \AA, He\,{\sc i}  5876 \AA, H$\alpha$ 6563 \AA, and He\,{\sc i} 6678 \AA. Within each of their respective orders, we crop out a small section of the spectrum encompassing the absorption feature, fit a polynomial to the surrounding continuum, divide by the best-fitting polynomial to normalize the spectrum, and fit a Gaussian to the absorption feature. We use the centroid of the best-fitting Gaussian as the observed wavelength in order to derive a velocity. Each spectrum is assigned a final radial velocity/uncertainty using a weighted average/uncertainty from all five individual line results. Finally, we convert these measurements to heliocentric velocities using PyAstronomy's {\em baryCorr} function. A sine wave fit to the data reveals a velocity semi-amplitude of $K = 200.6\pm2.3$\,km s$^{-1}$, as shown in Figure \ref{fig:EVR001_RV_fig}, with all radial velocity data provided in Table \ref{tab:rv_points} in the appendix. However, our individual exposure times were non-negligible fractions of the orbital period ($\sim$7.1\%). Orbital phase smearing leads to our measuring only 0.9917 of the full semi-amplitude (derivation shown in \citealt{bal99}); thus, we should inflate our measurement by a factor of 1.0084 to recover the true value. We report as our final semi-amplitude for the hot subdwarf primary $K = 202.3\pm2.3$ km s$^{-1}$. Additionally, we report a systemic velocity of $\gamma = 18.4\pm1.5$\,km s$^{-1}$ for the binary.

\begin{figure}[h]
\includegraphics[width=1\columnwidth]{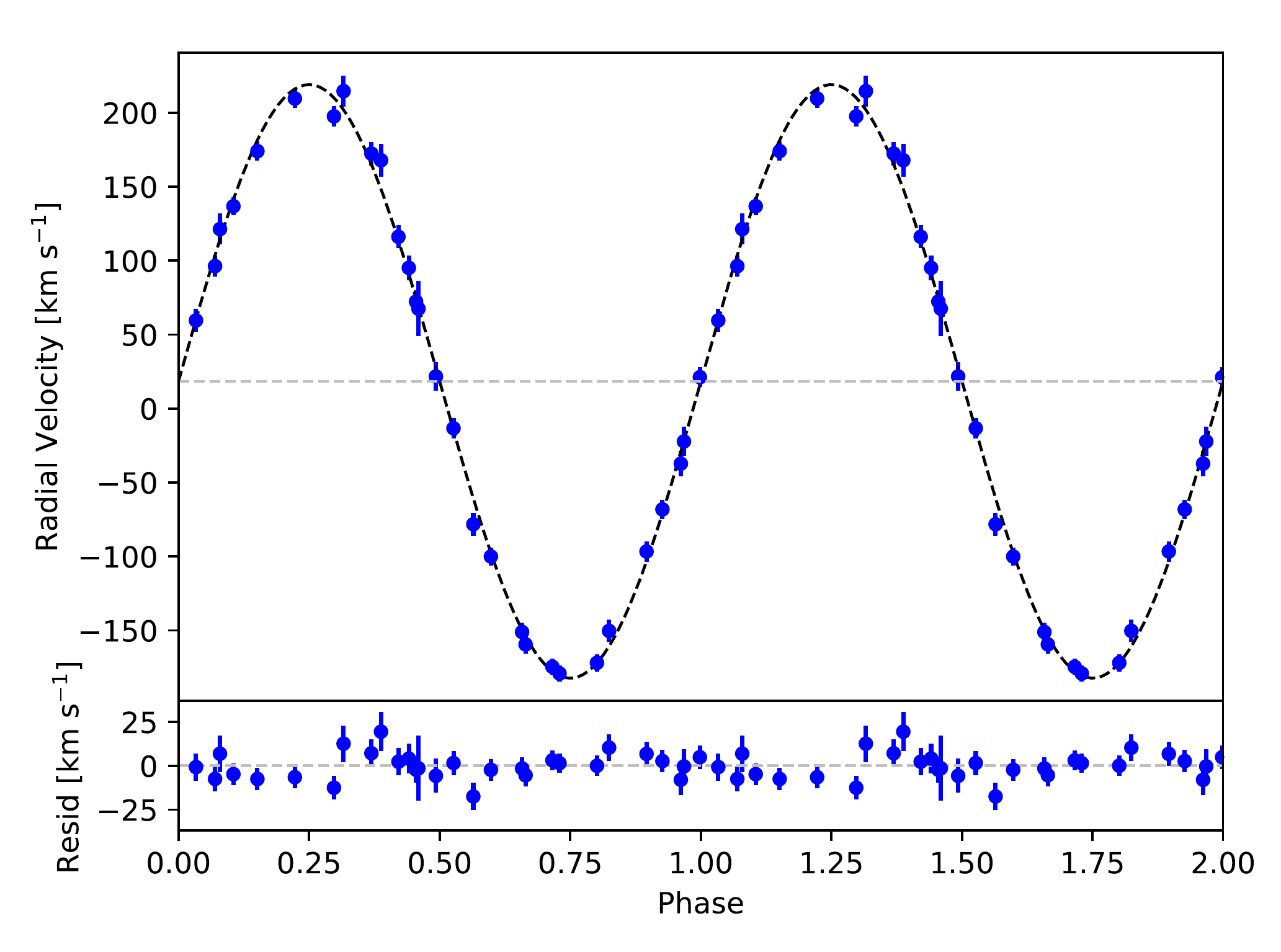}
\caption{{\em Top panel:} Phase-folded, heliocentric radial velocity measurements from SMARTS 1.5-m/CHIRON, plotted twice for better visualization. The solid line denotes the best-fitting sine wave to the data. After correcting for slight phase smearing, we find a velocity semi-amplitude of $K = 202.3\pm2.3$ km s$^{-1}$ and systemic velocity of $\gamma = 18.4\pm1.5$ km s$^{-1}$. {\em Bottom panel:} Residuals after subtracting the best-fitting sine wave from the data.}
\label{fig:EVR001_RV_fig}
\end{figure}

We use the rest-wavelength-corrected average SOAR spectrum to determine the primary star's atmospheric parameters by a simultaneous fitting of H and He line profiles with metal-line-blanketed LTE synthetic spectra, as described in \cite{2000A&A...363..198H}. The primary star's surface gravity ($\log(g)$), effective temperature ($T_{\mathrm{eff}}$), and helium abundance ($\log(y)=\log[n_{\rm He}/n_{\rm H}]$) are determined by fitting H Balmer profiles H13 through H$\beta$, along with He\,{\sc i} 4472 \AA. We note that the Balmer lines closest to the Balmer jump are the most sensitive to $\log(g)$ and $T_{\mathrm{eff}}$. We find $T_{\mathrm{eff}} = 18500\pm500$ K, $\log(g) = 4.96\pm0.04$, and $\log(y) = -1.34\pm0.11$. Errors were derived using a $\chi$-squared minimization. 
While the high-resolution CHIRON spectra are not suitable for determining $T_{\mathrm{eff}}$ and $\log(g)$, due to the H Balmer lines spanning multiple orders, they are sufficient for measuring the projected rotational velocity $v_{\mathrm{rot}} \sin i$ and more precisely determining the He abundance. After Doppler-correcting all CHIRON spectra to the same rest frame and stacking them to create a master high-resolution spectrum, we fitted the same synthetic models to the data, this time fixing $T_{\mathrm{eff}}$ and $\log(g)$ to the values determined from the SOAR spectrum. We find a helium abundance of $\log(y) = -1.43\pm0.03$, in agreement with the SOAR/Goodman result, along with a rotational velocity of  $v_{\mathrm{rot}} \sin i = 112\pm4$ km s$^{-1}$.

All final results from the atmospheric modeling are shown in Table \ref{solutions_table}. The derived parameters place the primary star in EVR-CB-001 at the extreme cool edge of known hot subdwarf B stars.

\begin{figure}[t]
\centering
\includegraphics[width=0.9\columnwidth]{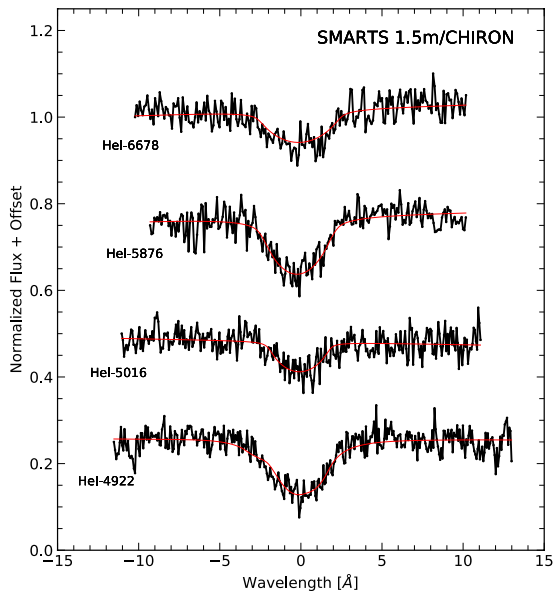}
\caption{Normalized SMARTS 1.5-m/CHIRON spectrum of EVR-CB-001 (black line) with best--fitting atmospheric model (red line). Parameters associated with the best-fitting LTE model spectrum are shown in the figure. $T_{\mathrm{eff}}$ and log $g$ were held as fixed parameters during the model fitting, set to the values determined from the SOAR/Goodman spectrum. }
\label{fig:spec_CHIRON}
\end{figure}


\section{Light Curve Analysis} \label{section_analysis_lc}

Since only spectral features from the primary star are detected, we must rely on light curve modeling to compute the mass ratio $q$ and constrain the system's parameters. We use the modeling code {\sc lcurve} \citep{2010MNRAS.402.1824C} to analyze both the SOAR and Evryscope light curves. {\sc lcurve} models the surface of each star using Roche lobe geometry and grids of points, and it takes into account gravity darkening, limb darkening, Doppler boosting, and mutual illumination effects. In order to constrain the parameter space searched by the models, we use several assumptions, boundary conditions, and results from spectroscopy. We assume the orbit is circular, and that the primary star's rotation is synchronized with the orbit. For the invisible companion we assume a lower limit to the radius (mass), using the zero-temperature mass-radius relation by Eggleton (quoted from \citealt{verbunt88}). The limb darkening prescription and the passband specific gravity darkening prescription was used following \citet{cla04,blo11} and as tabulated in \citet{claret11}. For the gravity darkening we used $b = 0.41\pm0.03$ for V and $b = 0.40\pm0.03$ for $g^\prime$. For limb darkening we used $a_1 = 0.76, a_2 = -0.18, a_3 = 0.10, a_4 = -0.03$ for V band and $a_1 = 0.71, a_2 = -0.27, a_3 = 0.17, a_4 = -0.05$ for $g^\prime$. Using the results for surface gravity ($\log{g}$), effective temperature ($T_{\mathrm{eff}}$), and rotational velocity ($v_{\mathrm{rot}}\sin(i)$) from \S~\ref{section_analysis_spectra} as a prior, combined with the orbital period ($P$) and radial velocity ($K$), we determine the inclination angle ($i$), the mass ratio ($q$), as well as the scaled radii and velocity scale (($K_{1}+K_{2})/\sin{i}$). Additionally we used a third order polynomial to account for residual airmass effects in the SOAR lightcurve. The subscript 1 is used for the object which dominates the light ($K_{1},M_{1},R_{1}$), and the subscript 2 is used for the invisible companion ($K_{2},M_{2},R_{2}$).

This solution requires the additional assumptions of a lower limit He WD radius and fixed limb darkening coefficients explained in detail in \citep{Kupfer_2017}. The assumptions regarding the unseen companion suggest that it does not contribute substantially to the light curve. We test our assumptions by comparing the luminosity contributions of the primary and secondary for a range of likely radii and temperatures for the He WD companion. Using conservative estimates of 0.03 $R_{\odot}$ and $T_{eff}$ of 10,000 K for the He WD companion, the luminosity contribution is 0.5\%. This increases to 2.5\% if the He WD companion has an effective temperature of 20,000 K. In a test run of our solution, we included the He WD effective temperature and radius as free parameters and found that both were unconstrained in the model fits. Because the luminosity contribution is very small and the He WD fit is unconstrained, we have assumed a fixed $T_{eff}$ of 6000 K and a fixed radius of 0.02 $R_{\odot}$ for the He WD companion which implies a negligible luminosity contribution of ~0.1\%. The overall result of our solution (\S~\ref{section_system_solution}) did not change with this assumption.

We combine {\sc lcurve} with the MCMC implementation {\sc emcee} \citep{2013PASP..125..306F} to explore the parameter space, converge on a solution, and to determine the uncertainties. We used 512 chains and let them run for 2500 trials well beyond a stable solution was reached. The corner plot of the final solution is shown in Figure \ref{fig:mcmc_results} in the appendix.

\begin{figure}[h]
\includegraphics[width=1.0\columnwidth]{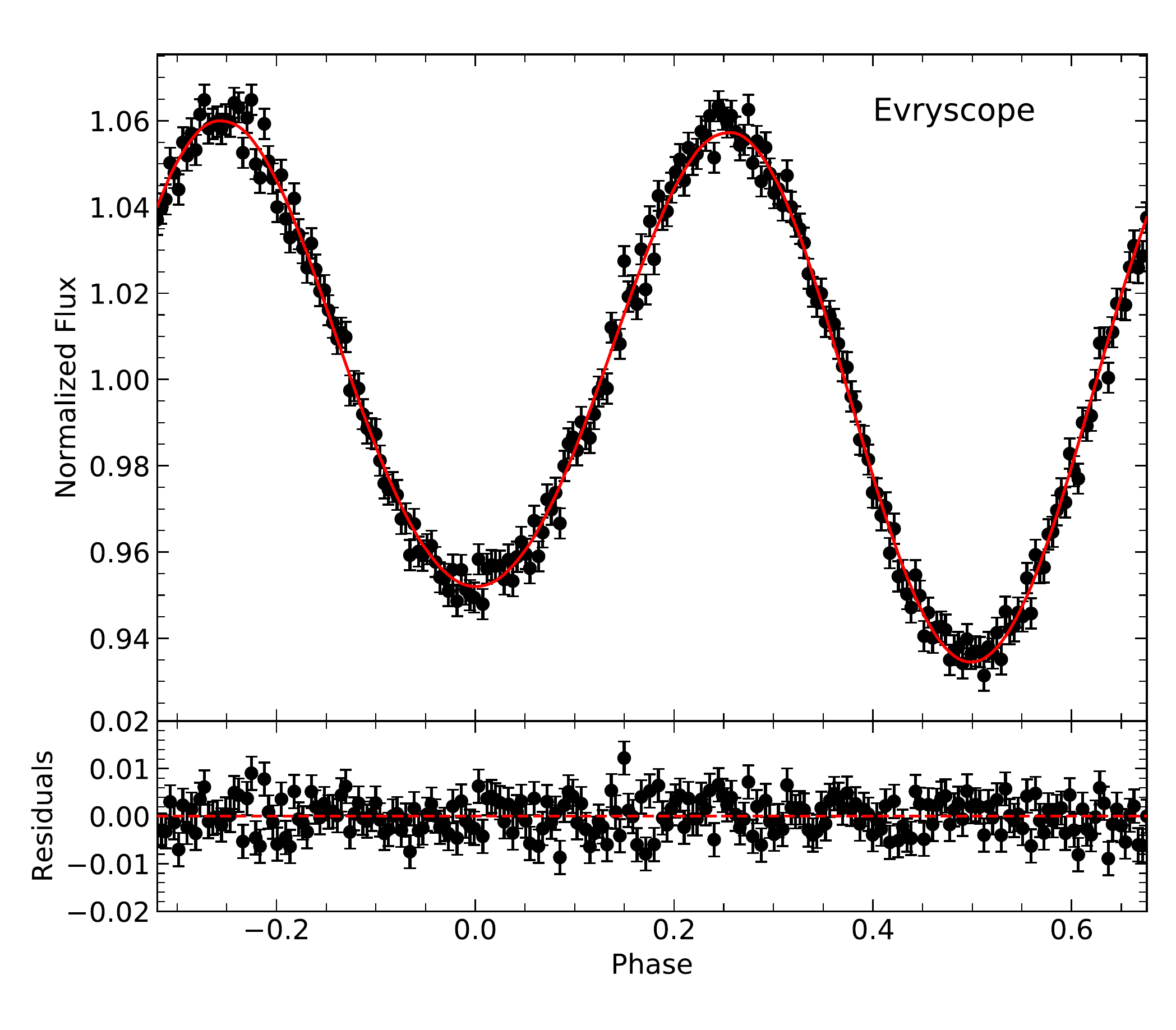}
\caption{{\em Top panel:} The binned in phase Evryscope $g$ light curve phase-folded on the 2.34252168 hour period with the best-fitting model determined by {\sc lcurve}. The original light curve has 53,698 epochs, and is binned using the unbiased $\sqrt{\# Epochs}=232$ points. {\em Bottom panel:} Residuals after subtracting the best-fitting model.}
\label{fig:LC_Evryscope}
\end{figure}

\begin{figure}[h]
\includegraphics[width=1.0\columnwidth]{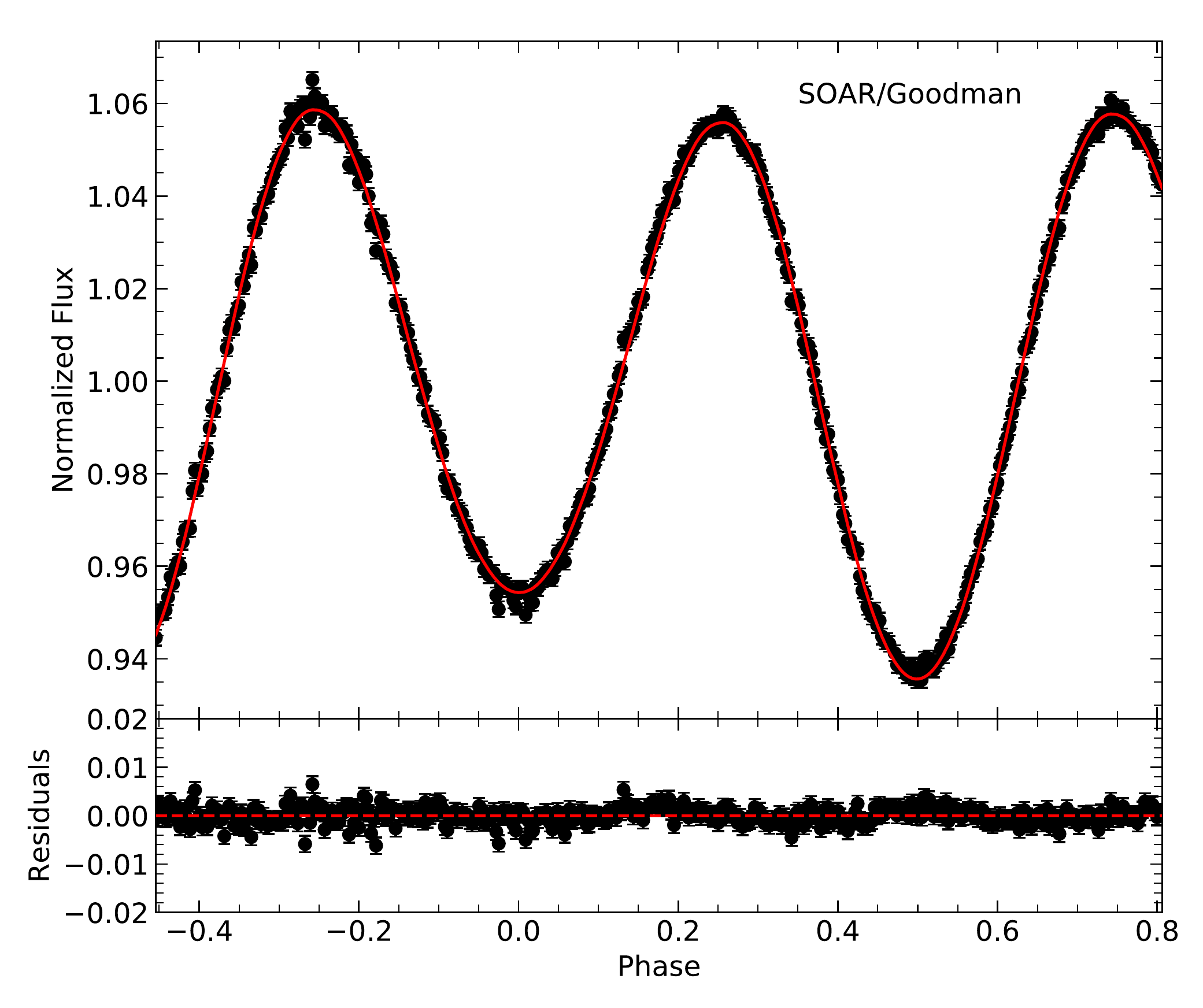}
\caption{{\em Top panel:} SOAR/Goodman $V$ light curve with the best-fitting model determined by {\sc lcurve}. {\em Bottom panel:} Residuals after subtracting the best-fitting model.}
\label{fig:LC_SOAR}
\end{figure}

We use the binary mass function

\begin{equation}
    f_{m} = \frac{M_{2}^3\text{sin}(i)^3}{(M_{1} + M_{2})^2} = \frac{PK^3}{2 \pi G}
\end{equation}
and assuming a tidally locked, circular orbit can be combined with
\begin{equation}
    \text{sin}(i) = \frac{(v_{rot} \text{sin}(i)) P}{2 \pi R_{1}}
\end{equation}
along with the standard mass-radius relation
\begin{equation}
    R_{1} = \sqrt{\frac{M_{1} G}{g}}
\end{equation}

to solve the system for the masses and radii of the visible ($M_{1},R_{1}$) and invisible component ($M_{2},R_{2}$). Full details of the approach are found in \citep{Kupfer_2017,gei07}. The final fits using the Evryscope binned in phase light curve is shown in Figure \ref{fig:LC_Evryscope} and the SOAR light curve is shown in Figure \ref{fig:LC_SOAR}. The ellipsoidal deformation dominates the photometric variation in the light curve, but Doppler boosting and gravity darkening effects are also present. We compare the Evryscope binned in phase light curve to the SOAR light curve in \S~\ref{section_es_advantage}.

\section{System Parameters} \label{section_system_solution}

EVR-CB-001 is a single-lined binary that does not show eclipses; consequently, we cannot determine a unique solution for the system from the light curve analysis alone. However, we can still constrain the masses and radii of the two stars by combining the results of the light curve modeling with results from the spectroscopic fitting and the assumption that the primary component is tidally synchronized with the orbit. Parameters derived in this way are summarized in Table \ref{solutions_table}.

Our solution converges on a mass ratio of $q$ = $M_1/M_2$ = $0.66\pm0.07$, with individual masses of $M_1$ = $0.21\pm0.05$ M$_{\odot}$ and $M_2$ = $0.32\pm0.06$ M$_{\odot}$. We reiterate that the lower-mass star of the two is the dominant source of light in the system, and the one showing ellipsoidal modulation. This object has a radius of $R_1$ =  $0.24\pm0.03$ R$_{\odot}$ showing that the low-mass primary star is a low-mass pre-WD. The radius ($R_2$) of the unseen companion cannot be determined, due to the lack of eclipses. However, since it does not produce any detectable light in the system despite its higher mass, the companion is consistent with a low-mass Helium white dwarf (He WD). 

\begin{table*}[ht]
\caption{Overview of Derived Parameters for EVR-CB-001}
\centering
\begin{tabular}{l l l l}
\hline
\hline
Description & Identifier & Units & Value\\
\hline
\multicolumn{4}{c}{Basic Information}\\
\hline
Evryscope ID & EVR-CB-001 &    &   \\
GAIA DR2 ID & 5216785445160303744 &    &   \\
Right ascension$^a$ & RA &  [deg]  & 132.06452462505\\
Declination$^a$ & Dec &  [deg]  & -74.31507593399\\
Magnitude$^a$ & $G$ & [mag] & 12.581$\pm$0.003\\
Parallax$^a$  & $\varpi$    & [mas] & 2.239$\pm$0.042 \\ 
Distance      &  d  & [pc] &   447$\pm$9  \\
Absolute Magnitude  &  M$_{\rm G}$     &   [mag]   & 4.33$\pm$0.05  \\ 
\hline
\multicolumn{4}{c}{Atmospheric Parameters of the Pre-He WD}\\
\hline
Effective temperature & $T_{\mathrm{eff}}$ & [K] & 18500$\pm$500\\
Surface gravity &  $\log(g)$ &    & 4.96$\pm$0.04\\
Helium abundance & $\log(y)$ &    & -1.43$\pm$0.03\\
Projected rotational velocity$^c$ & $v_{\mathrm{rot}} \sin i$ & [km $s^{-1}$] & 112$\pm$4\\
\hline
\multicolumn{4}{c}{Orbital Properties}\\
\hline
Period & P & [hr] & 2.3425217(5)\\
Reference phase$^b$ & $T_0$ & [BJD UTC] & 2457812.75378(5)\\
RV semi-amplitude & K & [km s$^{-1}$] & 202.3$\pm$2.3\\
Systemic velocity & $\gamma$ & [km $s^{-1}$] & 18.4$\pm$1.5\\
\hline
\multicolumn{4}{c}{Derived Parameters}\\\hline
Mass Ratio & $q$ &  & 0.66$\pm$0.07 \\
Pre-He WD mass & $M_{\mathrm{1}}$ & [$M_{\odot}$] & 0.21$\pm$0.05\\
Pre-He WD radius & $R_{1}$ & [$R_{\odot}$] & 0.24$\pm$0.03\\
He WD mass & $M_{2}$ & [$M_{\odot}$] & 0.32$\pm$0.06\\
Orbital inclination & $i$ & [$^\circ$] & 63$\pm$7\\
Separation & $a$ & [$R_{\odot}$] & 0.72$\pm$0.05 \\ 
\hline
\multicolumn{4}{l}{$^a${\em Gaia} G magnitude taken from the Gaia DR2 catalog \citep{GaiaDR2}}\\
\multicolumn{4}{l}{$^b$Time of light minimum, which corresponds to phase = 0.5 throughout the paper.}\\
\multicolumn{4}{l}{$^c$Slight phase smearing}
\end{tabular}
\label{solutions_table}
\end{table*}


\section{DISCUSSION} \label{section_discussion}

The primary component in EVR-CB-001 was originally thought to be a hot subdwarf B star, but the mass and surface gravity we have derived fall below the values of typical hot subdwarfs. Consequently, it is likely to be a post-RGB, pre-He WD, currently evolving through the cool end of the $T_{\mathrm{eff}}$-$\log(g)$ diagram occupied by hot subdwarfs. We independently estimate the mass of the pre-He WD and discuss its probable formation and evolution below.

\subsection{Independent mass estimate of the pre-He WD}

\subsubsection{Magnitude / Distance}

We tested our interpretation of the primary as a pre-He WD by estimating the pre-He WD radius and mass independently from the light curve modeling. Using the parallax from GAIA-DR2 \citep{GaiaDR2} we determine the distance, and with the Johnson V-band magnitude from APASS \citep{2015AAS...22533616H} we use the distance modulus to determine the absolute magnitude (with the bolometric and extinction corrections described below). With the mass-radius relation (equation 3), we express the luminosity ($L = 4 \sigma \pi R^2T^4$ from the Stephan-Boltzman equation applied to a black body) as a function of mass and surface gravity instead of radius. Using the zero-point luminosity, we solve for the mass, combine constants, and simplify to the following formula:

\begin{eqnarray*}
   M_{1}[M_{\rm odot}] &= 4.06609\times 10^{10}\times 10^{\log{g}} \times T_{\rm eff}[K]^{-4}  \\
   &\phantom{=} \quad \times 10^{-0.4\times(BC_{\rm V}+m_{\rm V}-A_{\rm V}+5\times \log{\varpi[arcsec])}}
 \end{eqnarray*}

In addition to the previously derived values for $T_{\rm eff}$ and $\log{g}$ from \S~\ref{section_analysis_spectra}, and the Gaia parallax $\varpi$, we adopted the apparent magnitude in the Johnson V-band $m_{\rm V}=12.619\pm0.051\,{\rm mag}$ from the APASS catalog. To account for the significant variability of the star, we adopted a higher uncertainty of $0.12\,{\rm mag}$. The bolometric correction $BC_{\rm V}=-1.76\pm0.075$ was interpolated from Vizier table J/A+A/333/231/table3 \citep{1998A&A...333..231B} for the appropriate spectroscopic parameters. The extinction $A_{\rm V}=0.3007\pm0.027\,{\rm mag}$ was taken from the Stilism 3D maps of the local interstellar medium\footnote{https://stilism.obspm.fr/} \citep{2014A&A...561A..91L} adopting the parallax distance from GAIA. To derive the mass uncertainty we used the Python Monte Carlo error propagation \textit{mcerp} package  assuming that all input parameters are normally distributed. From the resulting distribution we adopted the maximum value and the mass values at FWHM to derive the uncertainties.

In this way we  derive $M_{1}=0.30_{-0.10}^{+0.17}\,M_{\rm \odot}$ consistent with the mass determination from the binary analysis and indicating a low-mass pre-He WD. Using the mass-radius relation the radius of the pre-He WD $R_{1}=0.30_{-0.07}^{+0.09}$ is derived to be slightly larger than from the light curve analysis, but still consistent within the uncertainties.

\subsubsection{MESA Stellar Evolution Code}\label{sec:mesa}

\begin{figure}[h]
\includegraphics[width=0.9\columnwidth]{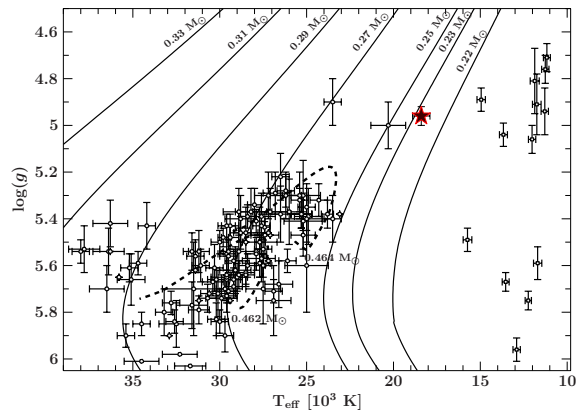}
\caption{MESA evolutionary tracks for a variety of pre-He WDs and low-mass He-burning star models. EVR-CB-001's atmospheric parameters are overplotted and show the primary star is likely a pre-He WD with mass near 0.2 M$_{\odot}$, in agreement with our light curve modeling solution. Known hot subdwarfs (open circles;  \citealt{2015A&A...576A..44K}) and some binaries from the ELM sample (open squares; \citealt{2016ApJ...818..155B}) are shown for comparison. EVR-CB-001 lies clearly in between the hot subdwarfs and the ELM sample.}
\label{fig:evo_path}
\end{figure}

To understand the nature of the primary, we have constructed pre-helium WD models for different masses using the MESA stellar evolution code \citep{pax11,pax13,pax15,pax18}, release version 10398. The models were constructed using an initially $1.0 \, M_\odot$ star (thought to be the most likely progenitor when starting at the main-sequence stage) that ascends the red-giant branch (RGB), building a helium core before it starts He-core burning. Once the helium core reaches a specified mass, all but $0.01\, M_\odot$ of the hydrogen envelope is stripped. Residual hydrogen shell burning then governs the timescale for evolution as the star contracts and evolves toward hotter $T_{\rm eff}$ as seen in the resulting tracks (solid lines) in Fig.\,\ref{fig:evo_path}. 

Additionally, we also computed MESA models of $0.461 \, M_\odot$ (understood to be the beginning of the helium burning stage post RGB) He-burning stars with two different hydrogen envelope masses: $1.0$ and $3.0 \times 10^{-3} \, M_\odot$. Our models use the MESA “predictive mixing” scheme for core convection to allow for proper growth of the convective He core and yield the correct core burning lifetime and luminosity \citep{pax18}. The tracks for these models extend from the beginning of core He burning through exhaustion of He in the core 150 Myr later. After this phase, the stars will begin He shell burning and evolve toward a hotter effective temperature. Our measured $T_{\mathrm{eff}}$ and $\log(g)$ intercepts tracks with masses in the range $0.22-0.23$ M$_{\odot}$, which is in agreement with the mass determined from our light curve modeling (see \S~\ref{section_system_solution}) as well as the determination using Gaia DR2 (see the previous paragraph). Known hot subdwarfs from \cite{2015A&A...576A..44K} are shown for comparison, clearly hotter and with higher surface gravity than EVR-CB-001. Overall, the compact binary EVR-CB-001 appears to contain a pre-He WD that is ellipsoidally deformed due the gravitational presence of an unseen He WD companion.

\subsection{Comparison to other Ellipsoidal Systems}

There does not appear to be an exact known analog for EVR-CB-001. In the following discussion, we compare the prominent features and components to known systems.

The photometric variability of EVR-CB-001 most resembles one of the exceptional massive WD / hot subdwarf compact binaries such as KPD 1930+2752, KPD 0422+5421, or CD-30 11223 \citep{2000MNRAS.317L..41M, 1998MNRAS.300..695K, 2041-8205-759-1-L25} but with a higher amplitude in light curve variability. This is reasonable given the lower mass and surface gravity as well as the bloated nature of the pre-He WD of EVR-CB-001 compared to hot subdwarfs. CSS 41177 is a rare eclipsing WD / WD compact binary with deep eclipses and relatively low mass WDs \citep{2015ASPC..493..313B}. However, both of the WDs are mature and there is no ellipsoidal deformation given the high surface gravity of each WD.

Short period WD / WD binaries with extremely low mass secondaries have been recently discovered showing tidal distortions \citep{2012ApJ...749...42H, 2014ApJ...792...39H} and eclipses in the case of the exceptional system J0651+2844 \citep{2012ApJ...757L..21H}. The ELM survey \citep{2011ApJ...727....3K} has discovered compact binaries and potential merger systems with extremely low mass secondary components. The higher temperature, lower surface gravity, early evolutionary stage, and extreme light curve variation of EVR-CB-001 are quite different compared to the ELM binaries, as is the mass ratio of the primary and secondary. 

The companion of HD 188112 \citep{2003A&A...411L.477H} is perhaps most similar to the pre-He WD of EVR-CB-001, however it is higher mass, surface gravity, and temperature. The system is quite different than EVR-CB-001 with a high mass WD primary, a longer period, and without photometric variation. WD 1242-105 \citep{2015AJ....149..176D} is an example double degenerate binary with similarly favorable conditions to EVR-CB-001 that will potentially merge into a single hot subdwarf B star. The higher total mass and similar primary and secondary WDs highlight some of the differences to the EVR-CB-001 system. OWJ074106.0-294811.0 \citep{2017ApJ...851...28K} is an ultra compact system with large photometric variability, but with quite different components (more massive, hotter, and higher surface gravity). EVR-CB-001 is best understood as combining interesting parts of each of these rare binaries to form a peculiar system.

\subsection{Formation History \& Future Evolution}

EVR-CB-001 likely formed via two separate stages of mass transfer. The original binary consisted of two main sequence stars. As the more massive of these evolved off the main sequence first and ascended the red giant branch, it filled its Roche lobe and started transferring mass onto its less massive companion. Whether this stage of mass transfer was dynamically stable (stable RLOF) or unstable (CE formation) depends on many unknown parameters, most notably the mass ratio at the time of transfer. Either way, enough mass was stripped from the red giant that its remnant was unable to fuse helium thereafter and formed a He WD. 

\begin{figure*}[t]
\includegraphics[width=2.1\columnwidth]{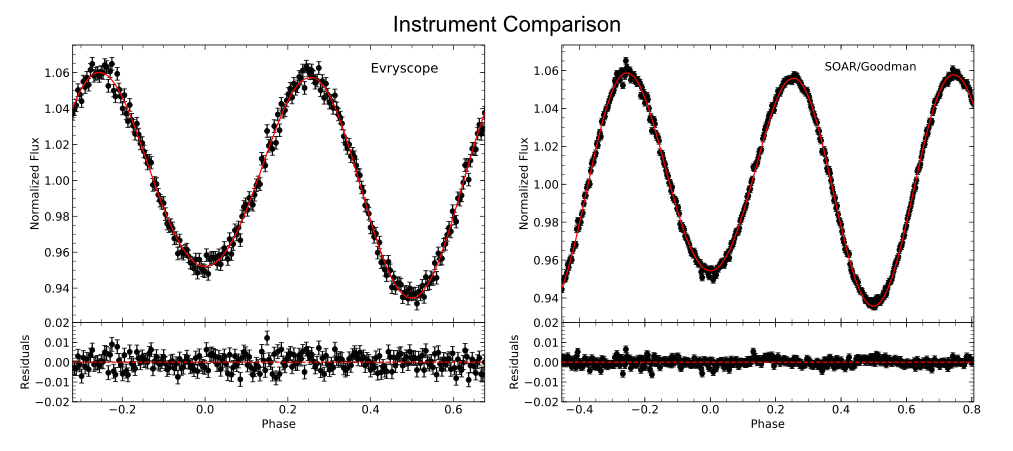}
\caption{Instrument comparison of the Evryscope and SOAR telescopes. \textit{Left Panel:} The Evryscope binned-in-phase light curve and the residuals after removing the best fit from \S~\ref{section_analysis_lc}. \textit{Right Panel:} The SOAR light curve and the residuals after removing the best fit from \S~\ref{section_analysis_lc}. The flux and residual scales are the same for both instruments to aid in the comparison. The Evryscope aperture is $\approx$ 4500 times smaller than SOAR, but produces a competitive light curve when binned-in-phase. This result is made possible by the improvement from combining the many period observations over the multi-year Evryscope survey time.}
\label{fig:rms_comp_fig}
\end{figure*}

The second phase of mass transfer, which commenced once the lower-mass main sequence star reached the giant phase, was undoubtedly unstable and led to the formation of a common envelope. Its He WD companion was unable to accrete at a sufficiently high rate, and significant mass was ejected, further tightening the orbit. Once again, the stripped object was left with insufficient mass for fusing He, causing it also to bypass the horizontal branch and collapse onto the white dwarf cooling sequence as another He WD. We appear to have caught EVR-CB-001 fairly shortly after this second mass transfer stage: the object that was most recently stripped of its outer layers appears as a hot and bloated pre-He WD, on its way to becoming a fully-degenerate WD. Assuming that the progenitor of the pre-He WD was a $\sim$1 M$_{\odot}$ star, we can calculate the orbital period of EVR-CB-001 at the moment when the progenitor filled its Roche lobe. Using the same MESA model as used in \S~\ref{sec:mesa} we find that a 1\,M$_{\odot}$ progenitor has a radius of 4 - 9\,R$_{\odot}$ when the helium core has built up a mass of 0.17 - 0.23\,M$_{\odot}$. Assuming a 0.3\,M$_{\odot}$ companion we find that the progenitor system consisting of a He-WD with a red giant had a period of $\approx$1 - 3 days when the red giant filled its Roche Lobe and started unstable mass transfer. This shows that the orbit must have shrunken substantially during the common envelope phase when the pre-He WD was formed. 

EVR-CB-001 represents a viable candidate progenitor system for the He WD merger channel leading to single hot subdwarf B stars (e.g. \citealt{han02, schwab18}). Eventually, the pre-He WD we now observe will evolve onto the white dwarf cooling sequence, and EVR-CB-001 will become a full-fledged double-degenerate system. At such a short orbital period, gravitational wave radiation will cause the system to shrink until the less massive He WD (currently a pre-He WD) fills its Roche lobe in about $\approx1$\,Gyr, at an orbital period of a few minutes. If the initiated mass transfer is dynamically unstable, the less massive He WD will be dynamically disrupted and form an accretion disk around its companion \citep{2012MNRAS.427..190S, 2012MNRAS.419..452Z}. Depending on the details of the evolution of the accretion disk and accretion rates, it is possible for the more massive He WD to increase its mass to the point where it ignites He shell burning and becomes a core He-burning hot subdwarf B star with $\sim$0.5 M$_{\odot}$. Unlike the other formation channels presented by \citet{han02,han03}, which all leave behind a {\em binary} hot subdwarf system, this He WD merger channel produces a {\em single} hot subdwarf B star. 

Although EVR-CB-001 is a candidate to form a single hot subdwarf B star, the system has a mass ratio (.66 $\pm$ .07) which might prevent the merger and instead evolve into a stable accreting AM\,CVn type binary. We briefly discuss this possibility here. For double white dwarf systems, commonly in the literature a system with a mass ratio $q=M_2/M_1<2/3$, $M_1$ being the mass of the accretor, is assumed to prevent the merger. However, \citet{mar04} and \citet{gok07} studied the effect of coupling of the accretor's and donor's spin to the orbit when the larger objects starts to fill its Roche Lobe. They found that a strong coupling and therefore a strong feedback of angular momentum to the orbit can destabilize systems with mass ratios lower than $q=M_2/M_1<2/3$,$M_1$ being the mass of the pre-He WD. Most recently, \citet{she15} proposed that even accreting double WD binaries with extreme mass ratios will merge due to classical nova-like outbursts on the accretor. Dynamical friction within the expanding nova shell causes the binary separation to shrink and the donor to dramatically overfill its Roche lobe, resulting in highly super-Eddington mass transfer rates that lead to a merger. This result was supported by \citet{bro16} who found that the merger rate of extremely low mass (ELM) white dwarfs exceeds the formation rate of AM\,CVn binaries by a factor of 40 concluding that most ELM white dwarf binaries merge. Thus, although we cannot definitively conclude either way, EVR-CB-001 is a viable candidate to merge and form a hot subdwarf B star.

\subsection{The Potential of the Evryscope } \label{section_es_advantage}

The Evryscope is a new instrument, different than a conventional telescope, and potentially misunderstood. Comparison of the Evryscope EVR-CB-001 discovery light curve to the SOAR followup light curve gives a powerful example of the Evryscope potential. Figure \ref{fig:rms_comp_fig} shows the binned-in-phase Evryscope light curve and SOAR light curve. The flux and residual scaling is the same in both plots. The astrophysical signal is fit with the best solution in \S~\ref{section_analysis_lc} and removed from both curves leaving the residuals. The residual RMS of the SOAR and Evryscope light curves is 0.00155 and 0.00354 respectively. Consider the following instrument comparisons: The Evryscope cameras are 6.1 cm diameter while the SOAR telescope is 4.1 meter diameter. The Evryscope instrument cost $\approx \$300K$, while the SOAR telescope cost $\approx \$28M$. The competitive Evryscope light curve is made possible because the SOAR light curve took 2.5 hours of observing time, while the Evryscope light curve took 2.5 years. SOAR observed 1 period, while the Evryscope observed over 1000. An individual Evryscope period observation has only a very modest precision (in this case $\approx$ .05 RMS), but with the proper photometric pipeline and systematics removal, the final combined and binned-in-phase light curve improves as $\approx \sqrt{\# periods}$ (in this case $\approx \sqrt{1000}$). 

It is important to emphasize that SOAR (or any other large telescope) and Evryscope are very different instruments. SOAR has many capabilities that Evryscope does not - spectroscopy, radial velocity measurements, and multi-band photometry just to name a few. It offers rapid precision followup on high value targets that the Evryscope cannot match. However, the Evryscope has a 8150 sq. deg. field of view with continuous 2-minute cadence that provides light curves just like the one for EVR-CB-001, but for 9.3M targets brighter than $m_v=15$. While some are better quality and some are worse depending on target brightness and location, EVR-CB-001 is a representative example. The Evryscope is a robotic system that requires minimal human intervention, with low construction and operating costs, and provides a dataset that facilitates the discovery of rare, difficult to detect, fast event systems like EVR-CB-001. With the proper processing of the discovery light curve, very high levels of binned-in-phase precision can be reached.


\section{SUMMARY} \label{section_summary}

We present the discovery of EVR-CB-001 - a close binary with an unseen low mass ($0.32 M_{\odot}$) helium white dwarf (He WD) and an extremely low mass progenitor helium white dwarf ($0.21 M_{\odot}$) (pre-He WD) companion. This object was discovered using Evryscope photometric data in a southern-all-sky hot subdwarf variability survey. EVR-CB-001 is a unique system: a short period (2.34 hours), large amplitude ellipsoidal modulation (12.0\% change in brightness from maximum to minimum) He WD / pre-He WD compact binary. Gravitational wave radiation will cause the system to shrink, and the helium rich WDs will potentially merge into a single hot subdwarf B star.

\section*{Acknowledgements}

This research was supported by the NSF CAREER grant AST-1555175 and the Research Corporation Scialog grants 23782 and 23822. HC is supported by the NSFGRF grant DGE-1144081. BB is supported by the NSF grant AST-1812874. The Evryscope was constructed under NSF/ATI grant AST-1407589. This work was supported by the National Science Foundation through grant PHY 17-148958. We acknowledge the use of the Center for Scientific Computing supported by the California NanoSystems Institute and the Materials Research Science and Engineering Center (MRSEC) at UC Santa Barbara through NSF DMR 1720256 and NSF CNS 1725797.

\clearpage
\bibliographystyle{apj}
\bibliography{hsd_wd_binary}

\begin{thebibliography}{}
\expandafter\ifx\csname natexlab\endcsname\relax\def\natexlab#1{#1}\fi

\bibitem[{{Baldry}(1999)}]{bal99}
{Baldry}, I. 1999, PhD thesis, University of Sydney

\bibitem[{{Bessell} {et~al.}(1998){Bessell}, {Castelli}, \&
  {Plez}}]{1998A&A...333..231B}
{Bessell}, M.~S., {Castelli}, F., \& {Plez}, B. 1998, \aap, 333, 231

\bibitem[{{Bloemen} {et~al.}(2011{\natexlab{a}}){Bloemen}, {Marsh},
  {{\O}stensen}, {Charpinet}, {Fontaine}, {Degroote}, {Heber}, {Kawaler},
  {Aerts}, \& {Green}}]{2011MNRAS.410.1787B}
{Bloemen}, S., {Marsh}, T.~R., {{\O}stensen}, R.~H., {et~al.}
  2011{\natexlab{a}}, \mnras, 410, 1787

\bibitem[{{Bloemen} {et~al.}(2011{\natexlab{b}}){Bloemen}, {Marsh},
  {{\O}stensen}, {Charpinet}, {Fontaine}, {Degroote}, {Heber}, {Kawaler},
  {Aerts}, {Green}, {Telting}, {Brassard}, {G{\"a}nsicke}, {Handler}, {Kurtz},
  {et~al.}}]{blo11}
---. 2011{\natexlab{b}}, MNRAS, 410, 1787

\bibitem[{{Bours} {et~al.}(2015){Bours}, {Marsh}, {Parsons}, {Copperwheat},
  {Dhillon}, {Littlefair}, {Gaensicke}, {Gianninas}, \&
  {Tremblay}}]{2015ASPC..493..313B}
{Bours}, M.~C.~P., {Marsh}, T.~R., {Parsons}, S.~G., {et~al.} 2015, in
  Astronomical Society of the Pacific Conference Series, Vol. 493, 19th
  European Workshop on White Dwarfs, ed. P.~{Dufour}, P.~{Bergeron}, \&
  G.~{Fontaine}, 313

\bibitem[{{Brown} {et~al.}(2016{\natexlab{a}}){Brown}, {Gianninas}, {Kilic},
  {Kenyon}, \& {Allende Prieto}}]{2016ApJ...818..155B}
{Brown}, W.~R., {Gianninas}, A., {Kilic}, M., {Kenyon}, S.~J., \& {Allende
  Prieto}, C. 2016{\natexlab{a}}, \apj, 818, 155

\bibitem[{{Brown} {et~al.}(2010){Brown}, {Kilic}, {Allende Prieto}, \&
  {Kenyon}}]{2010ApJ...723.1072B}
{Brown}, W.~R., {Kilic}, M., {Allende Prieto}, C., \& {Kenyon}, S.~J. 2010,
  \apj, 723, 1072

\bibitem[{{Brown} {et~al.}(2016{\natexlab{b}}){Brown}, {Kilic}, {Kenyon}, \&
  {Gianninas}}]{bro16}
{Brown}, W.~R., {Kilic}, M., {Kenyon}, S.~J., \& {Gianninas}, A.
  2016{\natexlab{b}}, \apj, 824, 46

\bibitem[{{Claret}(2004)}]{cla04}
{Claret}, A. 2004, A\&A, 428, 1001

\bibitem[{{Claret} \& {Bloemen}(2011)}]{claret11}
{Claret}, A., \& {Bloemen}, S. 2011, \aap, 529, A75

\bibitem[{{Clemens} {et~al.}(2004){Clemens}, {Crain}, \&
  {Anderson}}]{2004SPIE.5492..331C}
{Clemens}, J.~C., {Crain}, J.~A., \& {Anderson}, R. 2004, in \procspie, Vol.
  5492, Ground-based Instrumentation for Astronomy, ed. A.~F.~M. {Moorwood} \&
  M.~{Iye}, 331--340

\bibitem[{{Copperwheat} {et~al.}(2010){Copperwheat}, {Marsh}, {Dhillon},
  {Littlefair}, {Hickman}, {G{\"a}nsicke}, \&
  {Southworth}}]{2010MNRAS.402.1824C}
{Copperwheat}, C.~M., {Marsh}, T.~R., {Dhillon}, V.~S., {et~al.} 2010, \mnras,
  402, 1824

\bibitem[{{Debes} {et~al.}(2015){Debes}, {Kilic}, {Tremblay},
  {L{\'o}pez-Morales}, {Anglada-Escude}, {Napiwotzki}, {Osip}, \&
  {Weinberger}}]{2015AJ....149..176D}
{Debes}, J.~H., {Kilic}, M., {Tremblay}, P.-E., {et~al.} 2015, \aj, 149, 176

\bibitem[{{Eastman} {et~al.}(2010){Eastman}, {Siverd}, \& {Gaudi}}]{Eastman10}
{Eastman}, J., {Siverd}, R., \& {Gaudi}, B.~S. 2010, \pasp, 122, 935

\bibitem[{{Foreman-Mackey} {et~al.}(2013){Foreman-Mackey}, {Hogg}, {Lang}, \&
  {Goodman}}]{2013PASP..125..306F}
{Foreman-Mackey}, D., {Hogg}, D.~W., {Lang}, D., \& {Goodman}, J. 2013,
  Publications of the Astronomical Society of the Pacific, 125, 306

\bibitem[{{Gaia Collaboration} \& {et al.}(2018)}]{GaiaDR2}
{Gaia Collaboration}, \& {et al.} 2018, \aap, 616, A1

\bibitem[{{Geier} {et~al.}(2007){Geier}, {Nesslinger}, {Heber}, {Przybilla},
  {Napiwotzki}, \& {Kudritzki}}]{gei07}
{Geier}, S., {Nesslinger}, S., {Heber}, U., {et~al.} 2007, \aap, 464, 299

\bibitem[{{Geier} {et~al.}(2013){Geier}, {Marsh}, {Wang}, {Dunlap}, {Barlow},
  {Schaffenroth}, {Chen}, {Irrgang}, {Maxted}, {Ziegerer}, {Kupfer},
  {Miszalski}, {Heber}, {Han}, {Shporer}, {Telting}, {G{\"a}nsicke},
  {{\O}stensen}, {O'Toole}, \& {Napiwotzki}}]{2013A&A...554A..54G}
{Geier}, S., {Marsh}, T.~R., {Wang}, B., {et~al.} 2013, \aap, 554, A54

\bibitem[{{Gianninas} {et~al.}(2014){Gianninas}, {Dufour}, {Kilic}, {Brown},
  {Bergeron}, \& {Hermes}}]{2014ApJ...794...35G}
{Gianninas}, A., {Dufour}, P., {Kilic}, M., {et~al.} 2014, \apj, 794, 35

\bibitem[{{Gokhale} {et~al.}(2007){Gokhale}, {Peng}, \& {Frank}}]{gok07}
{Gokhale}, V., {Peng}, X.~M., \& {Frank}, J. 2007, \apj, 655, 1010

\bibitem[{{Hallakoun} {et~al.}(2016){Hallakoun}, {Maoz}, {Kilic}, {Mazeh},
  {Gianninas}, {Agol}, {Bell}, {Bloemen}, {Brown}, {Debes}, {Faigler}, {Kull},
  {Kupfer}, {Loeb}, {Morris}, \& {Mullally}}]{2016MNRAS.458..845H}
{Hallakoun}, N., {Maoz}, D., {Kilic}, M., {et~al.} 2016, \mnras, 458, 845

\bibitem[{{Han} {et~al.}(2003){Han}, {Podsiadlowski}, {Maxted}, \&
  {Marsh}}]{han03}
{Han}, Z., {Podsiadlowski}, P., {Maxted}, P.~F.~L., \& {Marsh}, T.~R. 2003,
  \mnras, 341, 669

\bibitem[{{Han} {et~al.}(2002){Han}, {Podsiadlowski}, {Maxted}, {Marsh}, \&
  {Ivanova}}]{han02}
{Han}, Z., {Podsiadlowski}, P., {Maxted}, P.~F.~L., {Marsh}, T.~R., \&
  {Ivanova}, N. 2002, \mnras, 336, 449

\bibitem[{{Heber}(1986)}]{heb86}
{Heber}, U. 1986, \aap, 155, 33

\bibitem[{{Heber}(2009)}]{heb09}
---. 2009, \araa, 47, 211

\bibitem[{{Heber}(2016)}]{2016PASP..128h2001H}
---. 2016, Publications of the Astronomical Society of the Pacific, 128, 082001

\bibitem[{{Heber} {et~al.}(2003){Heber}, {Edelmann}, {Lisker}, \&
  {Napiwotzki}}]{2003A&A...411L.477H}
{Heber}, U., {Edelmann}, H., {Lisker}, T., \& {Napiwotzki}, R. 2003, \aap, 411,
  L477

\bibitem[{{Heber} {et~al.}(2000){Heber}, {Reid}, \&
  {Werner}}]{2000A&A...363..198H}
{Heber}, U., {Reid}, I.~N., \& {Werner}, K. 2000, \aap, 363, 198

\bibitem[{{Henden} {et~al.}(2015){Henden}, {Levine}, {Terrell}, \&
  {Welch}}]{2015AAS...22533616H}
{Henden}, A.~A., {Levine}, S., {Terrell}, D., \& {Welch}, D.~L. 2015, in
  American Astronomical Society Meeting Abstracts, 336.16

\bibitem[{{Hermes} {et~al.}(2012{\natexlab{a}}){Hermes}, {Kilic}, {Brown},
  {Montgomery}, \& {Winget}}]{2012ApJ...749...42H}
{Hermes}, J.~J., {Kilic}, M., {Brown}, W.~R., {Montgomery}, M.~H., \& {Winget},
  D.~E. 2012{\natexlab{a}}, \apj, 749, 42

\bibitem[{{Hermes} {et~al.}(2012{\natexlab{b}}){Hermes}, {Kilic}, {Brown},
  {Winget}, {Allende Prieto}, {Gianninas}, {Mukadam}, {Cabrera-Lavers}, \&
  {Kenyon}}]{2012ApJ...757L..21H}
{Hermes}, J.~J., {Kilic}, M., {Brown}, W.~R., {et~al.} 2012{\natexlab{b}},
  \apjl, 757, L21

\bibitem[{{Hermes} {et~al.}(2014){Hermes}, {Brown}, {Kilic}, {Gianninas},
  {Chote}, {Sullivan}, {Winget}, {Bell}, {Falcon}, {Winget}, {Mason},
  {Harrold}, \& {Montgomery}}]{2014ApJ...792...39H}
{Hermes}, J.~J., {Brown}, W.~R., {Kilic}, M., {et~al.} 2014, \apj, 792, 39

\bibitem[{{Istrate} {et~al.}(2016){Istrate}, {Marchant}, {Tauris}, {Langer},
  {Stancliffe}, \& {Grassitelli}}]{ist16}
{Istrate}, A.~G., {Marchant}, P., {Tauris}, T.~M., {et~al.} 2016, \aap, 595,
  A35

\bibitem[{{Jayasinghe} {et~al.}(2019){Jayasinghe}, {Stanek}, {Kochanek},
  {Shappee}, {Holoien}, {Thompson}, {Prieto}, {Dong}, {Pawlak}, {Pejcha},
  {Shields}, {Pojmanski}, {Otero}, {Hurst}, {Britt}, \&
  {Will}}]{2019MNRAS.485..961J}
{Jayasinghe}, T., {Stanek}, K.~Z., {Kochanek}, C.~S., {et~al.} 2019, \mnras,
  485, 961

\bibitem[{Jones {et~al.}(2001)Jones, Oliphant, Peterson, {et~al.}}]{scipy}
Jones, E., Oliphant, T., Peterson, P., {et~al.} 2001, {SciPy}: Open source
  scientific tools for {Python}, [Online]

\bibitem[{{Kilic} {et~al.}(2011){Kilic}, {Brown}, {Allende Prieto},
  {Ag{\"u}eros}, {Heinke}, \& {Kenyon}}]{2011ApJ...727....3K}
{Kilic}, M., {Brown}, W.~R., {Allende Prieto}, C., {et~al.} 2011, \apj, 727, 3

\bibitem[{{Kilic} {et~al.}(2012){Kilic}, {Brown}, {Allende Prieto}, {Kenyon},
  {Heinke}, {Ag{\"u}eros}, \& {Kleinman}}]{2012ApJ...751..141K}
---. 2012, \apj, 751, 141

\bibitem[{{Koen} {et~al.}(1998){Koen}, {Orosz}, \&
  {Wade}}]{1998MNRAS.300..695K}
{Koen}, C., {Orosz}, J.~A., \& {Wade}, R.~A. 1998, \mnras, 300, 695

\bibitem[{Kovacs {et~al.}(2002)Kovacs, Zucker, \& Mazeh}]{Kovacs:2002gn}
Kovacs, G., Zucker, S., \& Mazeh, T. 2002, Astron. Astrophys., 391, 369

\bibitem[{{Kupfer} {et~al.}(2017{\natexlab{a}}){Kupfer}, {van Roestel},
  {Geier}, \& {Marsh}}]{Kupfer_2017}
{Kupfer}, T., {van Roestel}, J., {Geier}, S., \& {Marsh}, T.
  2017{\natexlab{a}}, The Astrophysical Journal, 835, 131

\bibitem[{{Kupfer} {et~al.}(2015){Kupfer}, {Geier}, {Heber}, {{\O}stensen},
  {Barlow}, {Maxted}, {Heuser}, {Schaffenroth}, \&
  {G{\"a}nsicke}}]{2015A&A...576A..44K}
{Kupfer}, T., {Geier}, S., {Heber}, U., {et~al.} 2015, \aap, 576, A44

\bibitem[{{Kupfer} {et~al.}(2017{\natexlab{b}}){Kupfer}, {Ramsay}, {van
  Roestel}, {Brooks}, {MacFarlane}, {Toma}, {Groot}, {Woudt}, {Bildsten},
  {Marsh}, {Green}, {Breedt}, {Kilkenny}, {Freudenthal}, {Geier}, {Heber},
  {Bagnulo}, {Blagorodnova}, {Buckley}, {Dhillon}, {Kulkarni}, {Lunnan}, \&
  {Prince}}]{2017ApJ...851...28K}
{Kupfer}, T., {Ramsay}, G., {van Roestel}, J., {et~al.} 2017{\natexlab{b}},
  \apj, 851, 28

\bibitem[{{Lallement} {et~al.}(2014){Lallement}, {Vergely}, {Valette},
  {Puspitarini}, {Eyer}, \& {Casagrande}}]{2014A&A...561A..91L}
{Lallement}, R., {Vergely}, J.~L., {Valette}, B., {et~al.} 2014, \aap, 561, A91

\bibitem[{Law {et~al.}(2015)Law, Fors, Ratzloff, Wulfken, Kavanaugh, Sitar,
  Pruett, {Birchard}, {Barlow}, {Cannon}, {Cenko}, {Dunlap}, {Kraus}, \&
  {Maccarone}}]{2015PASP..127..234L}
Law, N.~M., Fors, O., Ratzloff, J., {et~al.} 2015, PASP, 127

\bibitem[{{Lomb}(1975)}]{1975Ap&SS..39..447L}
{Lomb}, N.~R. 1975, Astrophysics and Space Science, 39, 447

\bibitem[{{Marsh} {et~al.}(1995){Marsh}, {Dhillon}, \&
  {Duck}}]{1995MNRAS.275..828M}
{Marsh}, T.~R., {Dhillon}, V.~S., \& {Duck}, S.~R. 1995, \mnras, 275, 828

\bibitem[{{Marsh} {et~al.}(2004){Marsh}, {Nelemans}, \& {Steeghs}}]{mar04}
{Marsh}, T.~R., {Nelemans}, G., \& {Steeghs}, D. 2004, \mnras, 350, 113

\bibitem[{{Maxted} {et~al.}(2000){Maxted}, {Marsh}, \&
  {North}}]{2000MNRAS.317L..41M}
{Maxted}, P.~F.~L., {Marsh}, T.~R., \& {North}, R.~C. 2000, \mnras, 317, L41

\bibitem[{{Nelemans} {et~al.}(2005){Nelemans}, {Napiwotzki}, {Karl}, {Marsh},
  {Voss}, {Roelofs}, {Izzard}, {Montgomery}, {Reerink}, {Christlieb}, \&
  {Reimers}}]{2005A&A...440.1087N}
{Nelemans}, G., {Napiwotzki}, R., {Karl}, C., {et~al.} 2005, \aap, 440, 1087

\bibitem[{{Ofir}(2014)}]{2014A&A...561A.138O}
{Ofir}, A. 2014, \aap, 561, A138

\bibitem[{{Paxton} {et~al.}(2011){Paxton}, {Bildsten}, {Dotter}, {Herwig},
  {Lesaffre}, \& {Timmes}}]{pax11}
{Paxton}, B., {Bildsten}, L., {Dotter}, A., {et~al.} 2011, ApJs, 192, 3

\bibitem[{{Paxton} {et~al.}(2013){Paxton}, {Cantiello}, {Arras}, {Bildsten},
  {Brown}, {Dotter}, {Mankovich}, {Montgomery}, {Stello}, {Timmes}, \&
  {Townsend}}]{pax13}
{Paxton}, B., {Cantiello}, M., {Arras}, P., {et~al.} 2013, ApJs, 208, 4

\bibitem[{{Paxton} {et~al.}(2015){Paxton}, {Marchant}, {Schwab}, {Bauer},
  {Bildsten}, {Cantiello}, {Dessart}, {Farmer}, {Hu}, {Langer}, {Townsend},
  {Townsley}, \& {Timmes}}]{pax15}
{Paxton}, B., {Marchant}, P., {Schwab}, J., {et~al.} 2015, ApJs, 220, 15

\bibitem[{{Paxton} {et~al.}(2018){Paxton}, {Schwab}, {Bauer}, {Bildsten},
  {Blinnikov}, {Duffell}, {Farmer}, {Goldberg}, {Marchant}, {Sorokina},
  {Thoul}, {Townsend}, \& {Timmes}}]{pax18}
{Paxton}, B., {Schwab}, J., {Bauer}, E.~B., {et~al.} 2018, \apjs, 234, 34

\bibitem[{{Postnov} \& {Yungelson}(2014)}]{2014LRR....17....3P}
{Postnov}, K.~A., \& {Yungelson}, L.~R. 2014, Living Reviews in Relativity, 17,
  3

\bibitem[{{Ratzloff} {et~al.}(2019{\natexlab{a}}){Ratzloff}, {Law}, {Fors},
  {Corbett}, {Howard}, {del Ser}, \& {Haislip}}]{2019arXiv190411991R}
{Ratzloff}, J.~K., {Law}, N.~M., {Fors}, O., {et~al.} 2019{\natexlab{a}}, arXiv
  e-prints, arXiv:1904.11991

\bibitem[{{Ratzloff} {et~al.}(2019{\natexlab{b}}){Ratzloff}, {Corbett}, {Law},
  {Barlow}, {Glazier}, {Howard}, {Fors}, {del Ser}, \& {Trifonov}}]{polarpaper}
{Ratzloff}, J.~K., {Corbett}, H.~T., {Law}, N.~M., {et~al.} 2019{\natexlab{b}},
  arXiv e-prints, arXiv:1905.02738

\bibitem[{{Rebassa-Mansergas} {et~al.}(2011){Rebassa-Mansergas}, {Nebot
  G{\'o}mez-Mor{\'a}n}, {Schreiber}, {Girven}, \&
  {G{\"a}nsicke}}]{2011MNRAS.413.1121R}
{Rebassa-Mansergas}, A., {Nebot G{\'o}mez-Mor{\'a}n}, A., {Schreiber}, M.~R.,
  {Girven}, J., \& {G{\"a}nsicke}, B.~T. 2011, \mnras, 413, 1121

\bibitem[{{Scargle}(1982)}]{1982Ap&SS..263..835S}
{Scargle}, J.~D. 1982, Astrophysics and Space Science, 263, 835

\bibitem[{{Schwab}(2018)}]{schwab18}
{Schwab}, J. 2018, \mnras, 476, 5303

\bibitem[{{Schwab} {et~al.}(2012){Schwab}, {Shen}, {Quataert}, {Dan}, \&
  {Rosswog}}]{2012MNRAS.427..190S}
{Schwab}, J., {Shen}, K.~J., {Quataert}, E., {Dan}, M., \& {Rosswog}, S. 2012,
  \mnras, 427, 190

\bibitem[{{Shen}(2015)}]{she15}
{Shen}, K.~J. 2015, \apj, 805, L6

\bibitem[{{Tamuz} {et~al.}(2005){Tamuz}, {Mazeh}, \&
  {Zucker}}]{2005MNRAS.356.1466T}
{Tamuz}, O., {Mazeh}, T., \& {Zucker}, S. 2005, \mnras, 356, 1466

\bibitem[{{Tokovinin} {et~al.}(2013){Tokovinin}, {Fischer}, {Bonati},
  {Giguere}, {Moore}, {Schwab}, {Spronck}, \&
  {Szymkowiak}}]{2013PASP..125.1336T}
{Tokovinin}, A., {Fischer}, D.~A., {Bonati}, M., {et~al.} 2013, \pasp, 125,
  1336

\bibitem[{Vennes {et~al.}(2012)Vennes, Kawka, O'Toole, Németh, \&
  Burton}]{2041-8205-759-1-L25}
Vennes, S., Kawka, A., O'Toole, S.~J., Németh, P., \& Burton, D. 2012, The
  Astrophysical Journal Letters, 759, L25

\bibitem[{{Verbunt} \& {Rappaport}(1988)}]{verbunt88}
{Verbunt}, F., \& {Rappaport}, S. 1988, \apj, 332, 193

\bibitem[{{Wang} \& {Han}(2012)}]{2012NewAR..56..122W}
{Wang}, B., \& {Han}, Z. 2012, New Astronomy Reviews, 56, 122

\bibitem[{{Zhang} \& {Jeffery}(2012)}]{2012MNRAS.419..452Z}
{Zhang}, X., \& {Jeffery}, C.~S. 2012, \mnras, 419, 452

\end{thebibliography}


\clearpage

\appendix \label{section_appendix}

\begin{figure}[h]
\includegraphics[width=1.0\columnwidth]{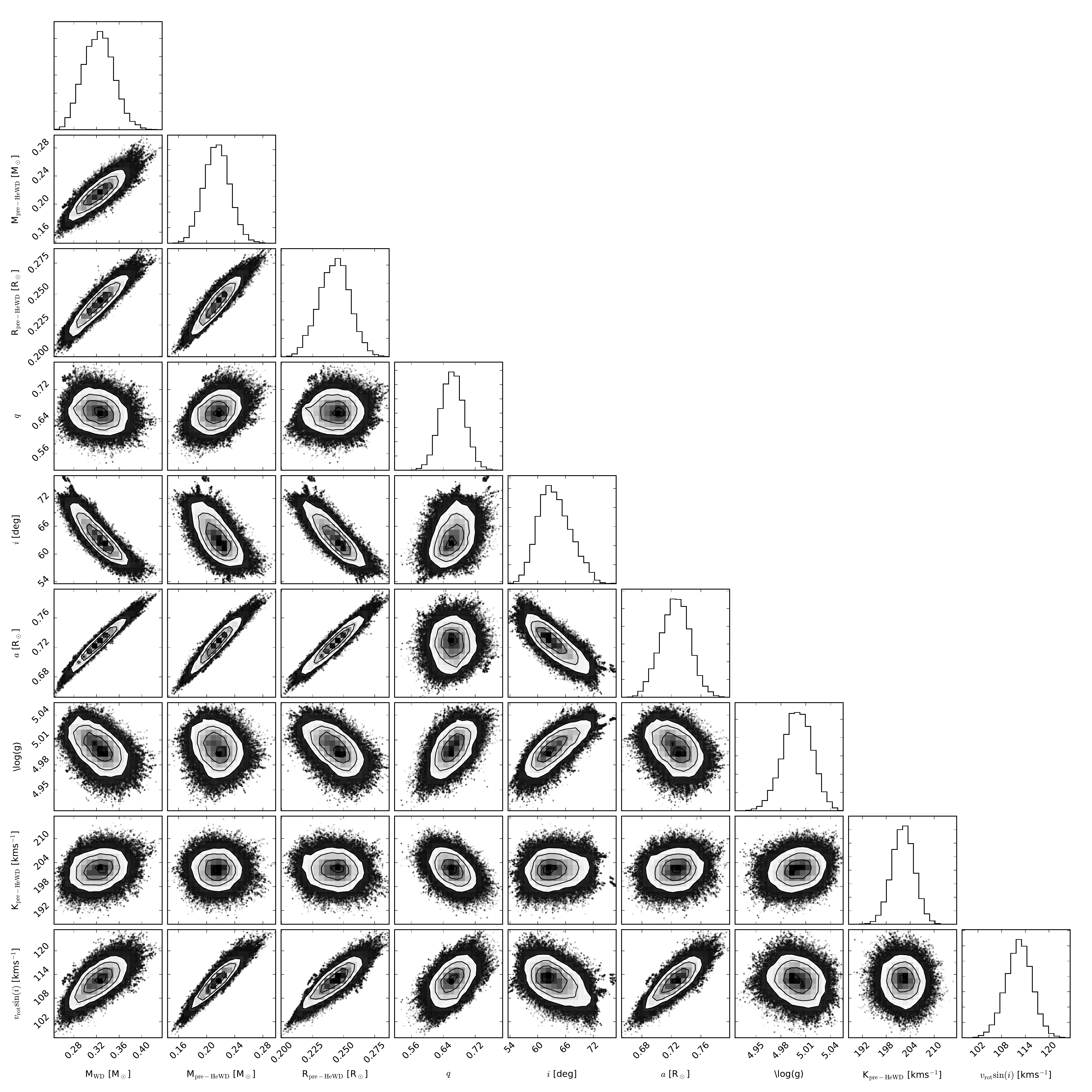}
\caption{Corner plots of the lightcurve fit of EVR-CB-001. The solution converged at low masses ($0.32 M_{\odot}$ for the He WD and $0.21 M_{\odot}$ for the pre-He WD), an inflated pre-He WD radius ($0.24 R_{\odot}$). Shown on the x-axis from left to right are: $M_{2}, M_{1}, R_{1}, i, a$ as well as the photometrically constrained $\log(g)$, velocity semi-amplitude $K_{1}$ and projected rotational velocity $v_{\rm rot}\sin(i)$.}
\label{fig:mcmc_results}
\end{figure}

\begin{figure*}[h]
\centering
\includegraphics[width=0.5\columnwidth]{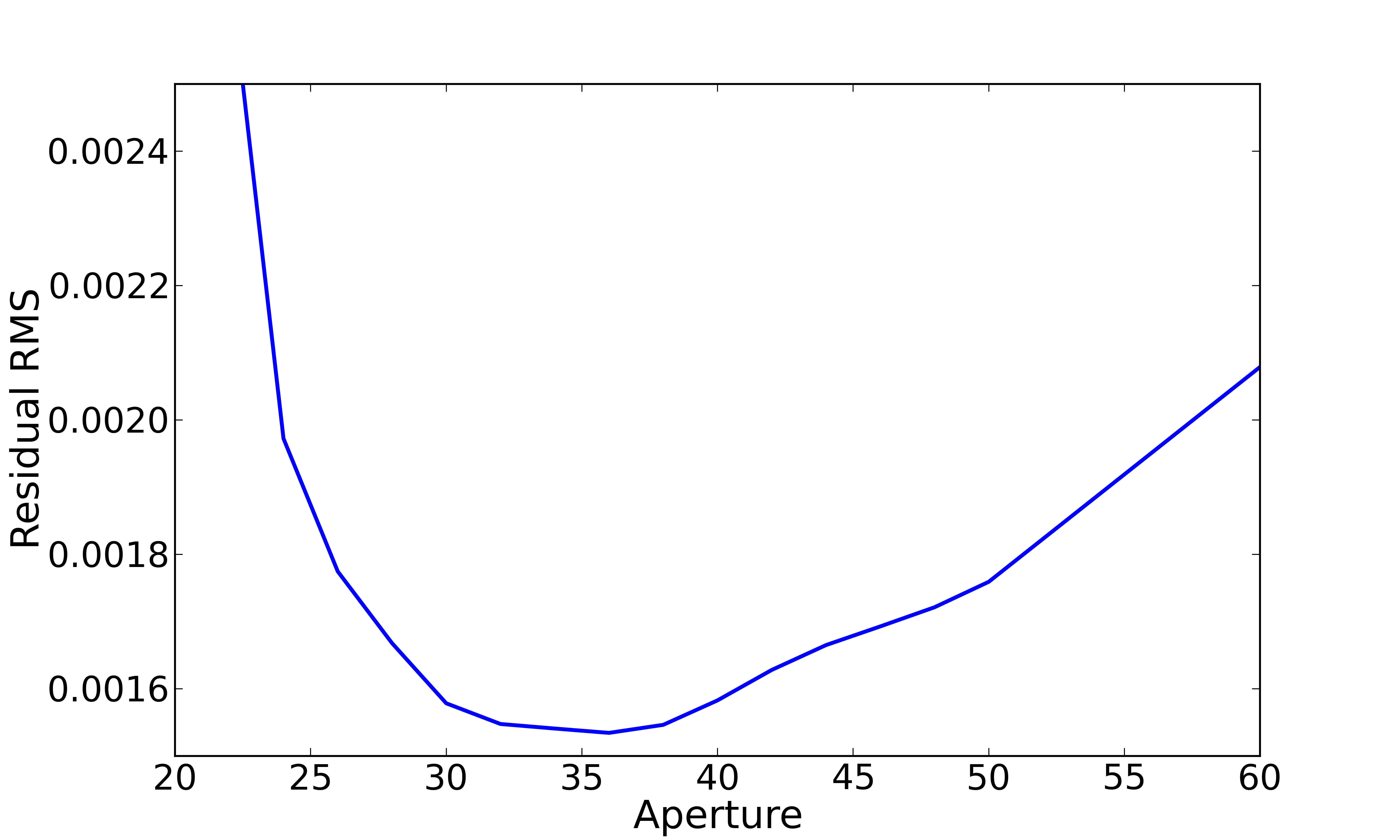}
\caption{The best aperture for the EVR-CB-001 SOAR light curve is a radius of 36 pixels giving a residual rms of .00153.}
\label{fig:rms_aperture}
\end{figure*}

\begin{table*}[ht]
\caption{Radial Velocity Measurements for EVR-CB-001}
\centering
\begin{tabular}{rrr}
\hline
Date & HRV & Error \\
(HJD - 2400000) & (km s$^{-1}$) & (km s$^{-1}$)\\
\hline
58471.803878 & -174.6 & 5.6\\
58472.774916 & -159.5 & 6.3\\
58480.807545 & -37.3 & 8.7\\
58480.814542 & 59.6 & 7.8\\
58480.821538 & 136.8 & 6.3\\
58481.780158 & -68.2 & 6.5\\
58481.787154 & 21.2 & 6.6\\
58481.794151 & 96.3 & 7.1\\
58483.747175 & 121.3 & 10.5\\
58483.754171 & 174.0 & 6.5\\
58483.761167 & 209.6 & 6.3\\
58484.759870 & 72.3 & 7.5\\
58484.766867 & -13.3 & 6.9\\
58484.773863 & -100.0 & 6.1\\
58494.797850 & 197.6 & 6.8\\
58494.804847 & 172.4 & 8.0\\
58494.811844 & 95.1 & 8.4\\
58495.678095 & 214.6 & 10.4\\
58495.685092 & 167.9 & 11.1\\
58495.692088 & 67.5 & 18.6\\
58498.616498 & 116.1 & 7.7\\
58498.623495 & 21.7 & 9.7\\
58498.630492 & -78.2 & 7.7\\
58508.806792 & -150.3 & 7.5\\
58508.813790 & -96.6 & 6.9\\
58508.820784 & -22.3 & 9.8\\
58511.718685 & -151.1 & 6.4\\
58511.725684 & -179.1 & 5.5\\
58511.732681 & -171.9 & 5.8\\
\hline
\end{tabular}
\label{tab:rv_points}
\end{table*}

\end{document}